\documentclass[pre,superscriptaddress,twocolumn]{revtex4-1}
\usepackage{amsfonts,amsmath,amssymb,mathtools,graphicx, float,blkarray, bigstrut}
\usepackage[colorlinks=true,allcolors=blue]{hyperref}
\usepackage[svgnames,dvipsnames,x11names]{xcolor}
\usepackage{tikz}
\usepackage{tikz-cd}
\usepackage{multirow}

\graphicspath{{Figures/}}

\def\ie{\rm{i.e.\ }}
\def\eg{\rm{e.g.\ }}
\newcommand\topstrut[1][1.2ex]{\setlength\bigstrutjot{#1}{\bigstrut[t]}}
\newcommand\botstrut[1][0.9ex]{\setlength\bigstrutjot{#1}{\bigstrut[b]}}

\begin{document}

\title{Course-Prerequisite Networks for Analyzing and Understanding Academic Curricula}

\author{Pavlos Stavrinides}
\affiliation{Department of Computing and Mathematical Sciences, California Institute of Technology, Pasadena, CA 91125, USA}

\author{Konstantin Zuev}
\email[Corresponding author, email: ]{kostia@caltech.edu}
\affiliation{Department of Computing and Mathematical Sciences, California Institute of Technology, Pasadena, CA 91125, USA}

\begin{abstract}
	
Understanding a complex system of relationships between courses is of great importance for the university's educational mission. This paper is dedicated to the study of course-prerequisite networks (CPNs), where nodes represent courses and directed links represent the formal prerequisite relationships between them. The main goal of CPNs is to model interactions between courses, represent the flow of knowledge in academic curricula, and serve as a key tool for visualizing, analyzing, and optimizing complex curricula. First, we consider several classical centrality measures, discuss their meaning in the context of CPNs, and use them for the identification of important courses. Next, we describe the hierarchical structure of a CPN using the topological stratification of the network. Finally, we perform the interdependence analysis, which allows to quantify the strength of knowledge flow between university divisions and helps to identify the most intradependent, influential, and interdisciplinary areas of study. We discuss how course-prerequisite networks can be used by students, faculty, and administrators for detecting important courses, improving existing and creating new courses, navigating complex curricula, allocating teaching resources, increasing interdisciplinary interactions between departments, revamping curricula, and enhancing the overall students' learning experience. The proposed methodology can be used for the analysis of any CPN, and it is illustrated with a network of courses taught at the California Institute of Technology. The network data analyzed in this paper is publicly available in the GitHub repository.  

\end{abstract}

% key words: course-prerequisite networks, curriculum analytics; curriculum design, complex curricula; network science;  

\maketitle

\section{Introduction}
\label{sec:Introduction}

An academic curriculum is a complex system of courses and interactions between them that lies at the heart of a university and underlies its educational mission. Understanding a university curriculum as a whole is an important prerequisite for providing students with a high quality education and meaningful learning experiences. Moreover, designing an appropriate curriculum is of great importance not only from an academic point of view, but also for organizational and financial management. 

A full list of courses together with their descriptions given in the university catalog allows, at least in principle, to know everything about the curriculum and answer any question about it. However, it is hard to comprehend this raw data, extract actionable knowledge, and make data-driven decisions. 

A network, where nodes represent courses and links represent certain relationships between them, is a natural model for conceptualizing, representing, and analyzing a curriculum. For example, links can represent the temporal relationships between courses based on how many students move from one course to another through out their studies~\cite{Akbas2015}. These temporal  networks can be used for forecasting course enrollments, predicting student performance~\cite{Slim2014}, and estimating the relative contribution of courses~\cite{Meghanathan2017}. Alternatively, links between courses can reflect the influence that some subjects have on others based on the expert knowledge of the professors. Such influence network models can be used for the curriculum design and recommendations~\cite{Blas2021}.

This paper focuses of the study of course-prerequisite networks (CPNs), where nodes represent courses and directed links represent the formal prerequisite requirements between them listed in the university catalog. Unlike temporal and influence networks, CPNs are objectively defined and, when in a steady state, don't change substantially from year to year. Over the last decade, CPNs have attracted a lot of interest from researchers due to their key role in the understanding of the complex structure of academic curricula. For example, Slim et al. used CPNs for detecting crucial courses that have a high impact on students progress and graduation rates~\cite{Slim2014a}, Aldrich discussed applications of CPNs to advising and curriculum reform~\cite{Aldrich2015}, and Molontay et al. introduced a data-driven probabilistic approach for studying the distribution of graduation time based on the CPN topology~\cite{Molontay2020}.

In this paper, we propose a general network-science-based framework for analysis of CPNs and illustrate it with a CPN based on the courses offered at the California Institute of Technology in the 2021-2022 academic year. We show that a CPN is an indispensable tool for visualizing, understanding, and optimizing an academic curriculum. It can be used not only for identification of important courses, but also for improving existing and creating new courses. We discuss how students can use a CPN to navigate their complex curriculum, and how a CPN can help faculty and administrators to meaningfully allocate teaching resources, increase interactions between divisions and departments, revamp the curriculum, and enhance the overall students' learning experience. 

The proposed framework is based on network science~\cite{Newman2018,NBW2006,Dorogovtsev2010,Easley2010}, which is an interdisciplinary field that emerged at the intersection of graph theory, computational statistics, computer science, and statistical physics. The basic idea of network science is to use a network as a  simplified representation of a complex system that captures the pattern of connection between system's components and represents its structural skeleton. Networks have been used to represent a variety of social, technological, information, and biological systems consisting of many interconnected, interacting components. Modeling complex systems with networks has proved to be useful for understanding systems as intricate and diverse as  the Internet, the world wide web, food webs, power grids, protein interactions, interwoven social groups, and even the human brain.

The rest of the paper is organized as follows. In Section~\ref{sec:NetworkRepresentation}, we define an abstract CPN and basic related notions and describe the Caltech CPN and its giant connected component. Section~\ref{sec:CentralityMeasures} is dedicated to the identification of important courses and different measures for importance quantification. In Section~\ref{sec:TopologicalOrdering}, we construct the topological stratification of a CPN and discuss how the emergent hierarchical structure on the CPN can be used for finding hidden prerequisites and creating comprehensive schedules for different areas of study. In Section~\ref{sec:InterdependenceAnalysis}, we perform an interdependence analysis of a CPN that provides a bird's eye view of the whole curriculum and allows to quantify the strength of flow of knowledge from one university division or area of study to another and identify the most intradependent, influential, and interdisciplinary divisions and areas of study. Finally, Section~\ref{sec:Conclusions} concludes with a brief summary and specific recommendations on how students, faculty, and administrators can use the results of the CPN analysis for efficient navigation and optimal enhancement of the curriculum.

\section{Network Representation of University Courses}
\label{sec:NetworkRepresentation}
The main object of study in this paper is a \textit{course-prerequisite network} (CPN), which is a directed network that describes interactions between university courses. In a CPN, nodes represent different courses and directed links between nodes represent the course-prerequisite relationships between the corresponding courses. A course $X$ is called a \textit{prerequisite} for course $Y$ if taking $X$ is required \textit{before} taking $Y$. Usually prerequisites cover material that is necessary for understanding more advanced courses. For example, a calculus course is often listed as a prerequisite for a course on differential equations. If $X$ is a prerequisite for $Y$, then, in the CPN, this is represented by a \textit{directed link} from node $X$ to node $Y$. In this case, $Y$ is called a \textit{postrequisite} of $X$. 

It is convenient to mathematically represent a CPN by its \textit{adjacency matrix}. If a CPN has $n$ nodes labeled by $1,\ldots,n$, then its adjacency matrix is the $n\times n$ matrix with elements $A_{ij}$, $i,j=1,\ldots,n$, defined as follows:
\begin{equation}
A_{ij}=\begin{cases}
1,& \mbox{if there is a link from } i \mbox{ to } j,\\
0,& \mbox{otherwise.}
\end{cases}
\end{equation}
The adjacency matrices of CPNs are \textit{sparse} (most of $A_{ij}$ equal to zero), since a typical course has a small number of prerequisites and serves as a prerequisite to a small number of courses.

\begin{figure}[h]
	\centering
	\includegraphics[width = \linewidth]{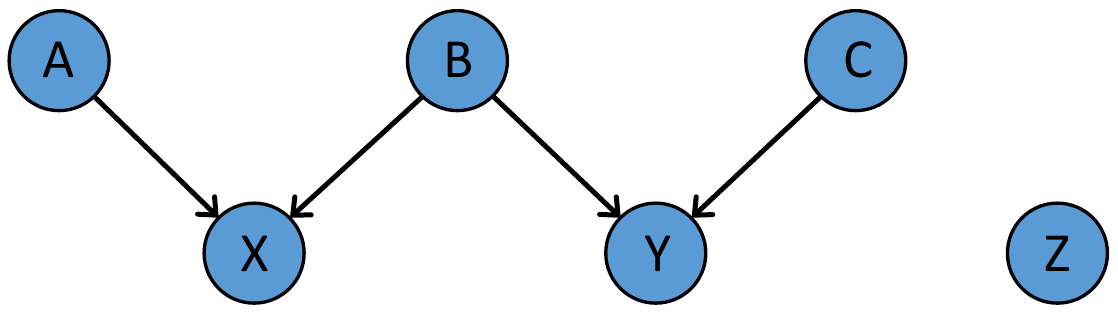}
	\caption{The course-prerequisite network (CPN) for the toy curriculum with six courses defined in Table~\ref{table:toy_example}.}
	\label{fig:toyCPN}
\end{figure}

As an example, consider a toy curriculum consisting of six courses: A, B, C, X, Y, and Z. The course-prerequisite relationships are summarized in Table~\ref{table:toy_example}.  
\begin{table}[h]
	\centering
	\begin{tabular}{c|c}
		\textsc{{Course}} & \textsc{{Prerequisites}}  \\
		\hline
		A & ---\\
		B & ---\\
		C & ---\\
		X & A, B\\
		Y & B, C\\
		Z & ---
	\end{tabular}
\caption{Example of a curriculum consisting of six courses.}
	\label{table:toy_example}
\end{table}
Course X has two prerequisites, A and B, course Y has two prerequisites, B and C, and all other courses have no prerequisites. If a course does not have any prerequisites, like courses A, B, C, and Z, then it can be taken any time. If a course does not have any prerequisites and postrequisites, like course Z, then it is represented by an \textit{isolated node}, \ie a node without incoming and outgoing links. Figure~\ref{fig:toyCPN} shows the CPN induced by the toy curriculum. 
%\begin{figure}[h]
%	\centering
%	\begin{tikzpicture}[node distance={18mm}, thick, main/.style = {draw, circle}]   
%	\node[main] (1) [minimum size=0.8cm, fill = blue!15!white] {A}; 
%	\node[main] (3) [minimum size=0.8cm, below right of=1, fill = blue!15!white] {X};
%	\node[main] (2) [minimum size=0.8cm, above right of = 3, fill = blue!15!white] {B};
%	\node[main] (4) [minimum size=0.8cm, below right of = 2, fill = blue!15!white] {Y};
%	\node[main] (5) [minimum size=0.8cm, above right of = 4, fill = blue!15!white] {C};
%	\node[main] (6) [minimum size=0.8cm, below right of = 5, fill = blue!15!white] {Z};
	
%	\draw (1) [->] to [out = 315, in = 135, looseness = 0.75] node[midway, above, sloped, pos=0.5]{} (3);
%	\draw (2) [->] to [out = 225, in = 45, looseness = 0.75] node[midway, above, sloped, pos=0.5]{} (3);
%	\draw (2) [->] to [out = 315, in = 135, looseness = 0.75] node[midway, above, sloped, pos=0.5]{} (4);
%	\draw (5) [->] to [out = 225, in = 45, looseness = 0.75] node[midway, above, sloped, pos=0.5]{} (4);
%	\end{tikzpicture}
%	\caption{The course-prerequisite network (CPN) for the toy curriculum with six courses defined in Table~\ref{table:toy_example}.}
%	\label{fig:toyCPN}
%\end{figure} 

The adjacency matrix of the this toy CPN is
\begin{equation}
A=
\begin{blockarray}{l c c c c c c}
&\text{A} & \text{B} & \text{C} & \text{X} & \text{Y} & \text{Z}\\
\begin{block}{l [c c c c c c]}
\text{A} & 0 & 0 & 0 & 1 & 0 & 0\topstrut\\
\text{B} &0 & 0 & 0 & 1 & 1 & 0\\
\text{C} &0 & 0 & 0 & 0 & 1 & 0\\
\text{X} &0 & 0 & 0 & 0 & 0 & 0\\
\text{Y} &0 & 0 & 0 & 0 & 0 & 0\\
\text{Z  } & 0 & 0 & 0 & 0 & 0 & 0\botstrut\\
\end{block}
\end{blockarray}
\end{equation}

A course-prerequisite network represents the flow of knowledge between different courses in a university curriculum. The main goal of this paper is to show how CPNs can be used for visualization of complex curricula, drawing important observations and insights about the courses, and helping students to navigate and faculty and administrators to optimize their curricula.

\begin{figure}[h]
	\centering
	\includegraphics[width = \linewidth]{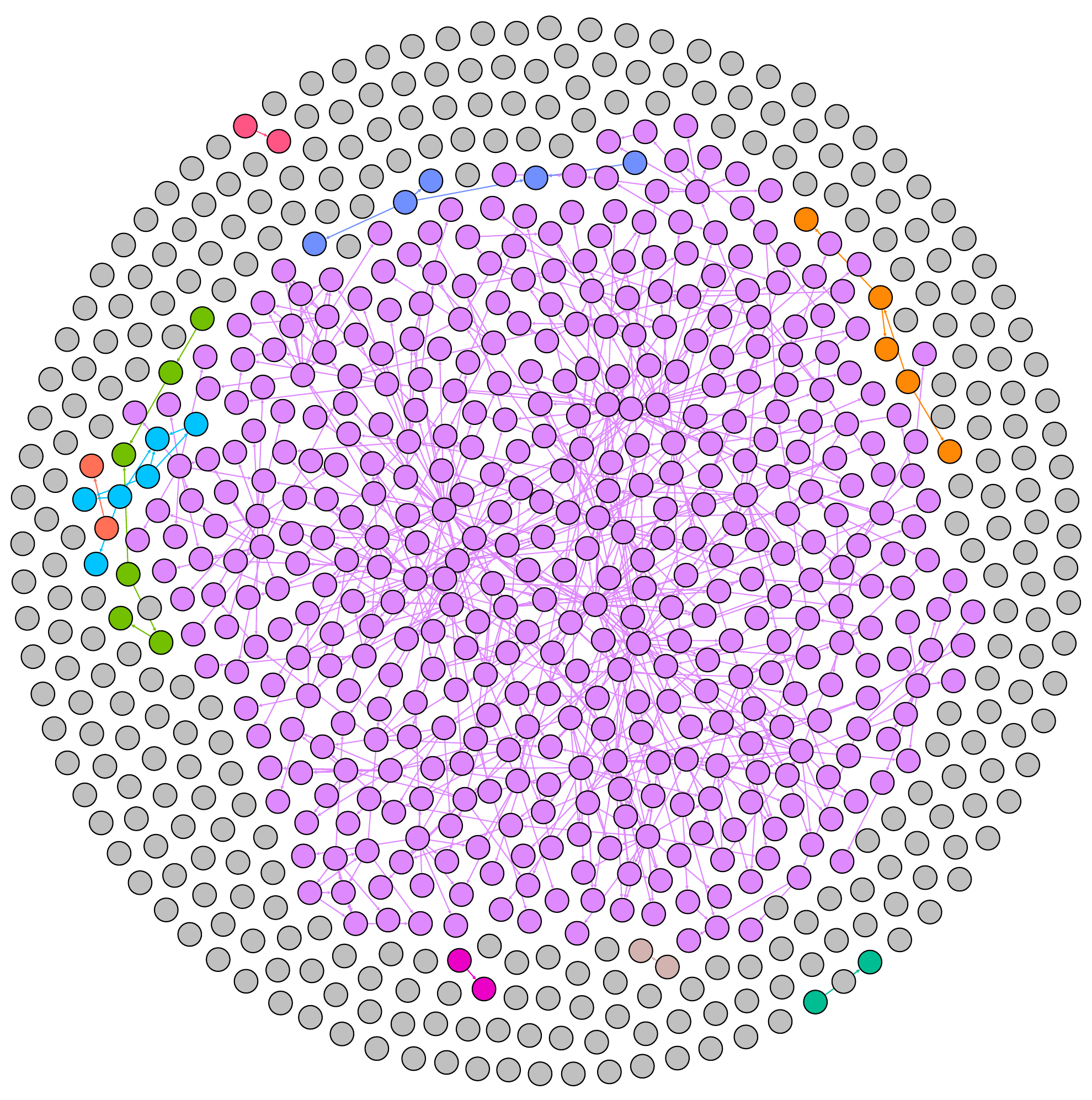}
	\caption{The 2021-2022 Caltech CPN. Nodes in gray are isolated, \ie have no prerequisites and do not serve as prerequisites. Nodes in color other than gray represent connected components. The network has $771$ nodes and $772$ links.}
	\label{fig:Caltech_CPN_full}
\end{figure}

The methods described here can be applied to the analysis of any CPN. We will illustrate them with a real CPN based on the courses that were offered at the \textit{California Institute of Technology} (Caltech) in the 2021-2022 academic year. The Caltech CPN, consisting of both undergraduate and graduate courses, is shown in Fig.~\ref{fig:Caltech_CPN_full}. All network visualizations in this paper are done in \textit{Gephi}, an open-source and free network visualization software package~\cite{Gephi}. For visual clarity, the network visualization in Fig.~\ref{fig:Caltech_CPN_full} omits the node labels (course names). A larger and more detailed visualization of the Caltech CPN is shown in Fig.~\ref{fig:Caltech_CPN_full_labels} in the Appendix. The network data is publicly available in the GitHub repository~\cite{GitHub}.  

Any university curriculum contains courses that are completely independent of each other: they are not prerequisites for each other, not prerequisites for another course, there is not a course which is a common prerequisite for them, etc. This independence between two courses manifests itself in the CPN by the absence of a \textit{path} between the nodes representing the courses. Independent courses belong to different \textit{connected components} of the CPN. Technically, a (weakly) connected component of a CPN is a subset of nodes such that for any two nodes in the subset there exists at least one path through the network connecting the nodes, where paths are allowed to go in both ways along any link (the directions of the links are ignored). Each isolated node constitutes a trivial connected component. Isolated nodes usually represent seminars, projects, outreach, and special topics courses. For example, the toy CPN on Fig.~\ref{fig:toyCPN} has two connected components: consisting of nodes \{A, B, C, X, Y\} and one isolated node Z.

Real-world CPNs  have several small connected components and one ``giant'' connected component, called the \textit{largest connected component} (LCC), which contains the largest fraction of nodes, almost all links, and constitutes the most interesting and nontrivial part of a CPN. The LCC of a CPN, denoted by $\mathcal{G}$, is the main part of the network, which represents its complex structure and function.   

In the Caltech CPN, in addition to isolated nodes, there are $10$ connected components represented by different colors in Fig.~\ref{fig:Caltech_CPN_full} and, in more detail, in Fig.~\ref{fig:Caltech_CPN_full_labels}.  All but one are very small, with sizes not exceeding six nodes. The largest connected component $\mathcal{G}$ (shown in pink) contains  $n=436$ nodes ($57$\% of all nodes) and  $m=747$ links ($97$\% of all links). In what follows, we focus our analysis on the largest connect component $\mathcal{G}$ of the Caltech CPN, which is shown in Fig.~\ref{fig:Caltech_LCC} in the Appendix.

\section{Centrality Measures}
\label{sec:CentralityMeasures}

One of the most interesting and intriguing questions about a university curriculum is the following: ``Which are the \textit{most important courses} in the curriculum?''. In other words, ``Which are the most important nodes in the CPN?''. Knowing the most important courses, the courses that form the ``backbone'' of the curriculum, could help to a) better allocate university resources to provide students with better experiences in these courses and b) inform students about these courses, so that they can pay special attention to them. 

There are different ways to define the ``importance'' of a node in a CPN. Here, we will consider three widely used in network science \textit{centrality measures}, which quantify the node importance: degree, PageRank centrality, and betweenness centrality.

\subsection{Degree Distributions}
\label{subsec:DegreeDistributions}

The \textit{total degree} of a node is the total number of links connected to it. In directed networks like CPNs, nodes have two kinds of degree: an \textit{in-degree}, the number of incoming links, and an \textit{out-degree}, the number of outgoing links. The in-degree $k_\text{in}(i)$ of a node $i$ is the number of prerequisites course $i$ has, and the out-degree $k_\text{out}(i)$ of $i$ is the number of courses for which $i$ is a prerequisite. In terms of the adjacency matrix $A$, the in- and out-degrees of node $i$ are given by
\begin{equation}
k_\text{in}(i)=\sum_{j=1}^nA_{ji} \hspace{5mm}\mbox{and}\hspace{5mm}k_\text{out}(i)=\sum_{j=1}^nA_{ij}.
\end{equation}
The total degree of node $i$ is then 
\begin{equation}
k(i)=k_\text{in}(i)+k_\text{out}(i)=\sum_{j=1}^n(A_{ij}+A_{ji}).
\end{equation}

The in-degree of a node measures how \textit{specialized} the corresponding course is: the larger $k_\text{in}(i)$ is, the more prerequisites course $i$ has, the more specialized it is. The out-degree, on the other hand, measures how \textit{fundamental} a course is:  the larger $k_\text{out}(i)$ is, the more courses have $i$ as a prerequisite, the more fundamental course $i$ is. We expect that in real-wold CPNs, the in- and out-degrees of nodes are negatively correlated. The absolute value of the Pearson correlation coefficient between in- and out-degrees of nodes,
\begin{equation}\hspace{-1mm}
\rho_\text{in,out}=\frac{\sum\limits_{i=1}^n\left(k_\text{in}(i)-\bar{k}_\text{in}\right)\left(k_\text{out}(i)-\bar{k}_\text{out}\right)}
{\sqrt{\sum\limits_{i=1}^n\left(k_\text{in}(i)-\bar{k}_\text{in}\right)^2}\sqrt{\sum\limits_{i=1}^n\left(k_\text{out}(i)-\bar{k}_\text{out}\right)^2}},
\end{equation} 
can be used to measure the structural difference between fundamental and specialized courses. 

\begin{figure}[h]
	\centering
	\includegraphics[width = 0.9\linewidth]{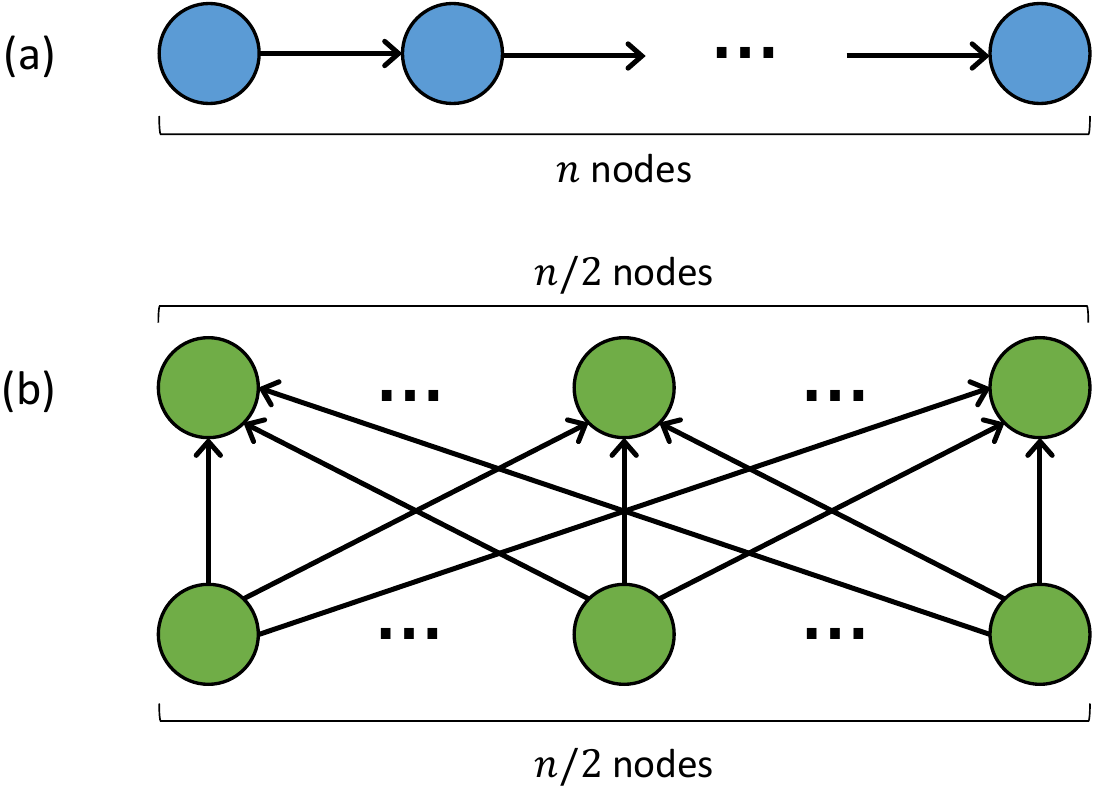}
	\caption{Two extreme cases. Curriculum (a) (blue): all courses are equivalent. Curriculum (b) (green): there is a substantial structural difference between fundamental courses (bottom row) and specialized courses (top row).}
	\label{fig:extreme_cases}
\end{figure}

To elaborate more on this, consider two extreme cases shown in Fig.~\ref{fig:extreme_cases}.

In curriculum (a), essentially all courses are structurally equivalent (except for the first and the last one); there are no fundamental and specialized courses. The sequences of in- and out-degrees are $k_\text{in}=(0,1,\ldots,1)$ and $k_\text{out}=(1,\ldots,1,0)$, and the correlation coefficient between them is
\begin{equation}
\rho_\text{in,out}^\text{(a)}=-\frac{1}{n-1}\approx 0, \hspace{5mm}\mbox{for large } n.
\end{equation}

In curriculum (b), however, there are fundamental courses (bottom row) and specialized courses (top row). These two types of courses are structurally very different: all fundamental courses are prerequisites for all specialized courses. The sequences of in- and out-degrees are $k_\text{in}=(0,\ldots,0,n/2,\ldots,n/2)$ and $k_\text{out}=(n/2,\ldots,n/2,0,\ldots,0)$,  and the correlation coefficient between them is
\begin{equation}
\rho_\text{in,out}^\text{(b)}=-1, \hspace{5mm}\mbox{for any } n.
\end{equation}

In real-world CPNs, the correlation coefficient $\rho_\text{in,out}\in(-1,0)$, and its magnitude $|\rho_\text{in,out}|$ quantifies the structural ``gap'' between fundamental and specialized courses: the larger $|\rho_\text{in,out}|$ is, the more significant the split of the curriculum into fundamental and specialized courses.

\begin{figure}[b]
	\centering
	\includegraphics[width = 0.8\linewidth]{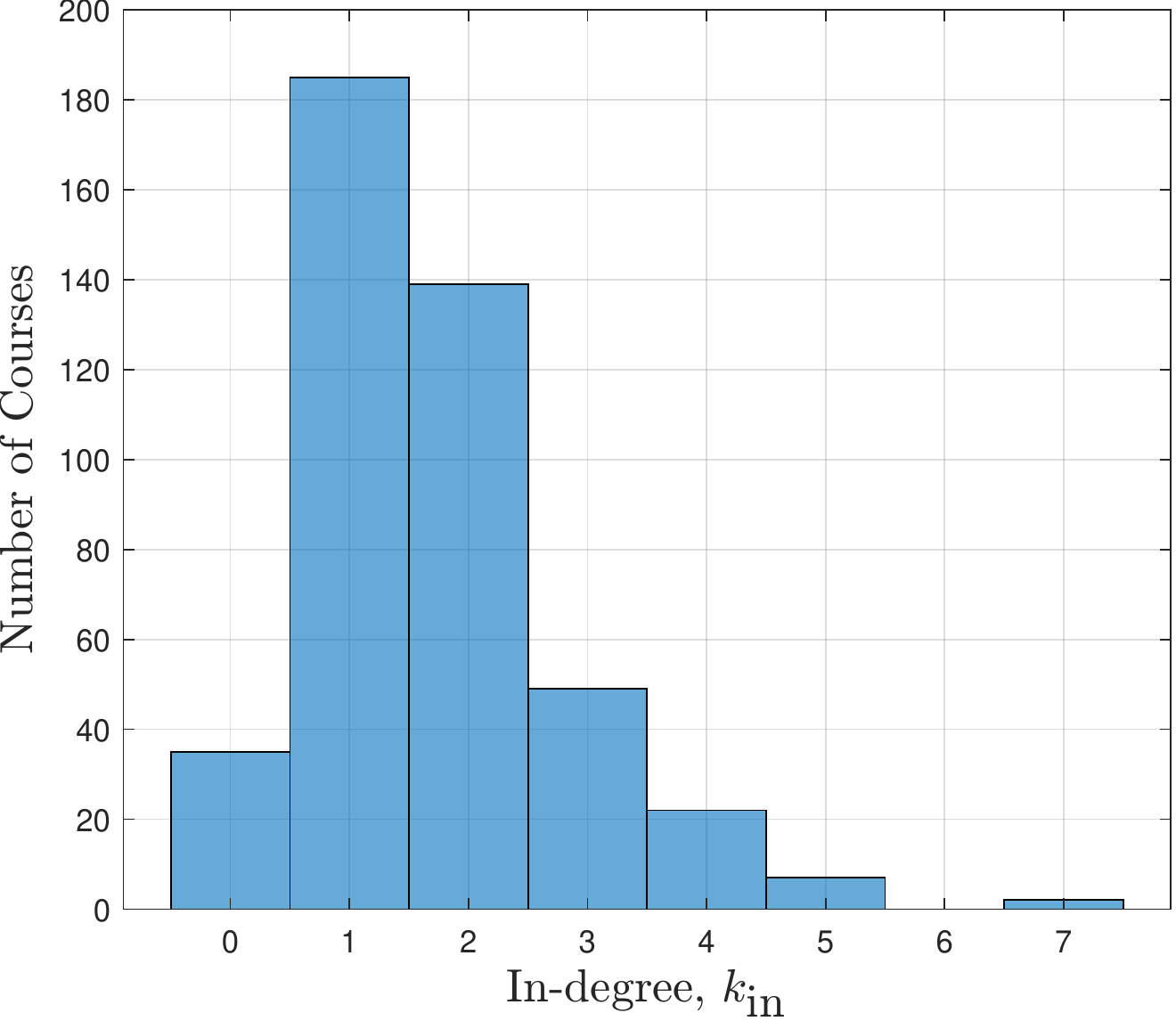}\\
	\includegraphics[width = 0.8\linewidth]{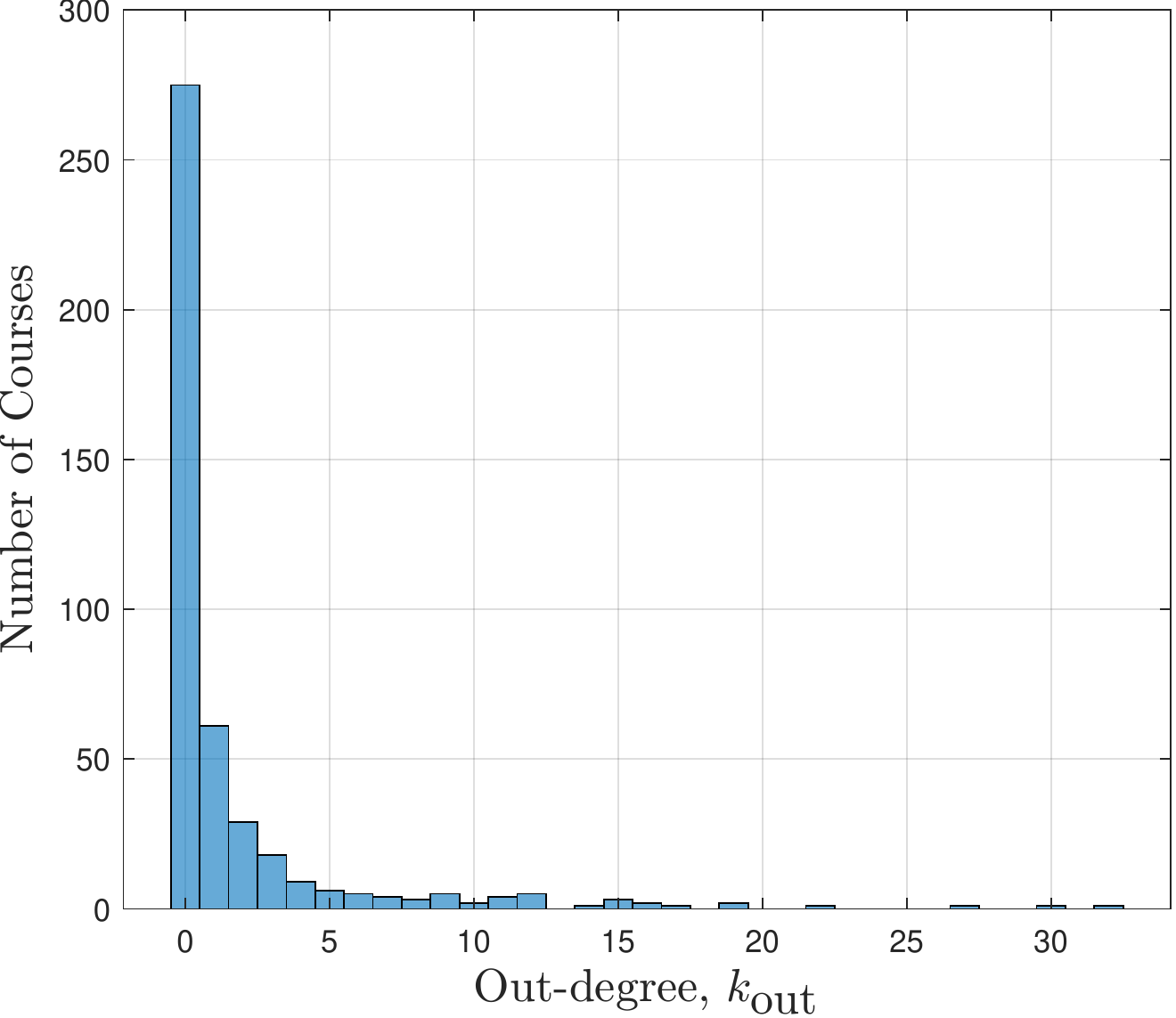}\\
	\includegraphics[width = 0.8\linewidth]{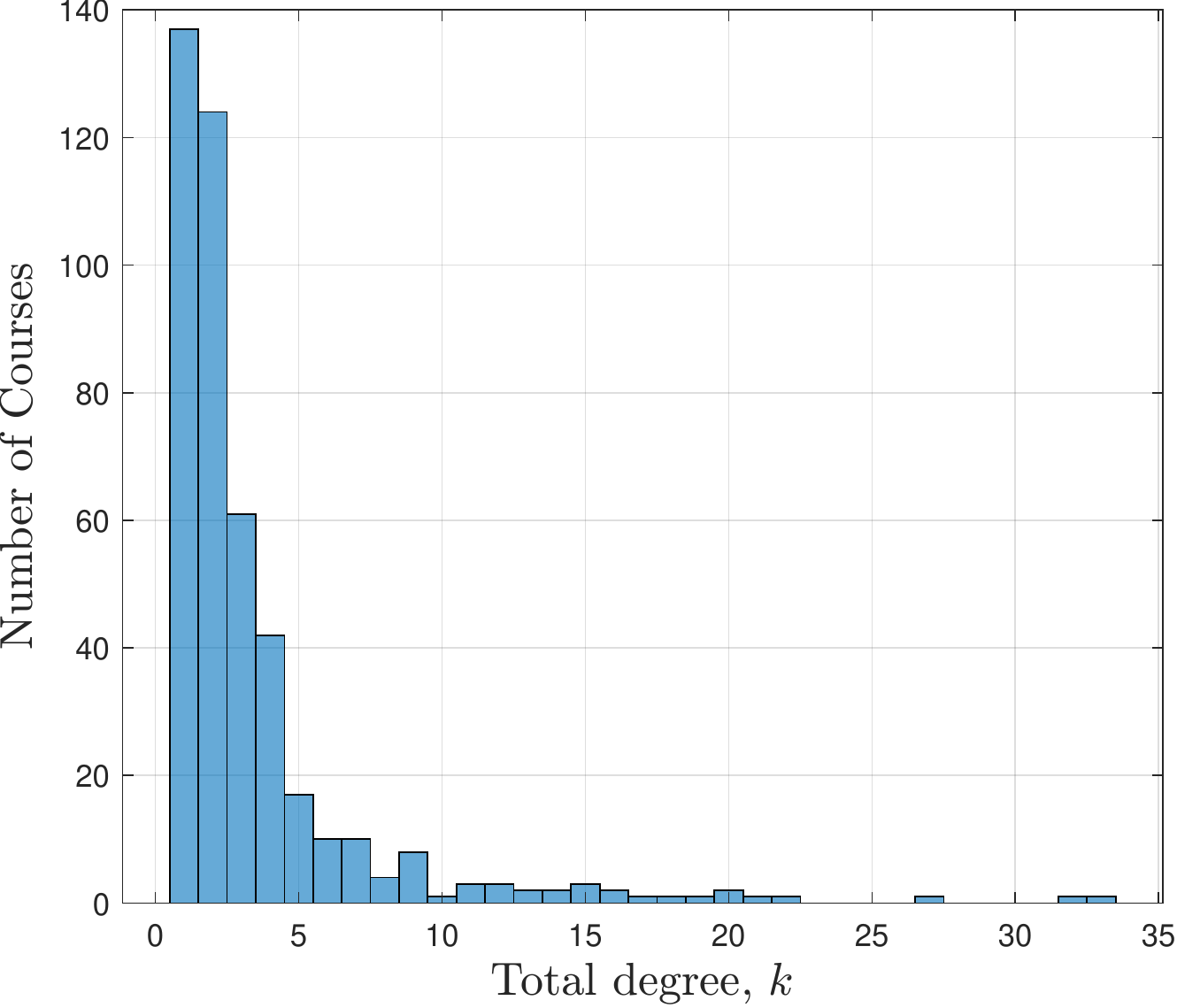}
	\caption{The in-, out-, and total degree distributions in the largest connected component $\mathcal{G}$ of the Caltech CPN.}
	\label{fig:deg_dists}
\end{figure}

Figure~\ref{fig:deg_dists} shows the histograms of in-, out-, and total degrees in the LCC $\mathcal{G}$ of the Caltech CPN. 
All degree distributions are right-skewed and have long right tails. 

A network is called \textit{scale-free}, a term coined in the seminal paper~\cite{BA1999}, if its degree distribution follows a power-law,
\begin{equation}
\mathbb{P}(k)\propto k^{-\gamma},\hspace{3mm}\textrm{for }k\geq k_{\text{min}},
\end{equation} 
where $\mathbb{P}(k)$ is the probability that a node chosen uniformly at random has degree $k$, $ k_{\text{min}}$ is the lower cut-off for the scaling region, and $\gamma$ is the power-law exponent. Many real-world networks are approximately scale-free, with the power-law exponent typically in the range $2<\gamma<3$~\cite{Newman2018}. For example, both the in- and out-degrees of the World Wide Web approximately follow power-law distributions  with $\gamma_{\text{in}}=2.1$ and $\gamma_{\text{out}}=2.7$~\cite{Broder2000}. For the LCC $\mathcal{G}$ of the Caltech CPN, the hypothesis test for discerning and quantifying power-law behavior in empirical data developed in \cite{CSN2009} accepts the hypothesis that the total degree distribution (bottom panel in Fig.~\ref{fig:deg_dists}) approximately follows a power-law, but rejects that hypothesis for the in- and out-degrees. The maximum-likelihood fitting method developed in~\cite{CSN2009} estimates the power-law exponent $\gamma$ and the lower cut-off $k_{\text{min}}$ as follows: 
\begin{equation}
\gamma=2.48 \hspace{3mm}\textrm{and}\hspace{3mm}k_{\text{min}}=3.
\end{equation} 

Most courses in $\mathcal{G}$ ($74$\%) have one or two prerequisites (top panel in Fig.~\ref{fig:deg_dists}) and many ($63$\%) are ``dead ends,'' \ie they do not serve as prerequisites for other courses (middle panel in Fig.~\ref{fig:deg_dists}). The average in-, out-, and total degrees are (note that always $\bar{k}_\text{in}=\bar{k}_\text{out}=\bar{k}/2$):
\begin{equation}
\begin{split}
\bar{k}_\text{in}&=\frac{1}{n}\sum_{i=1}^nk_\text{in}(i)=1.70, \\
\bar{k}_\text{out}&=\frac{1}{n}\sum_{i=1}^nk_\text{out}(i)=1.70,\\
\bar{k}&=\frac{1}{n}\sum_{i=1}^nk(i)=\bar{k}_\text{in}+\bar{k}_\text{out}=3.40.
\end{split}
\end{equation}

The Pearson correlation coefficient between in- and out-degrees of nodes is
\begin{equation}
\rho_\text{in,out}=-0.13.
\end{equation}

\begin{figure}[h]
	\centering
	\includegraphics[width = 0.8\linewidth]{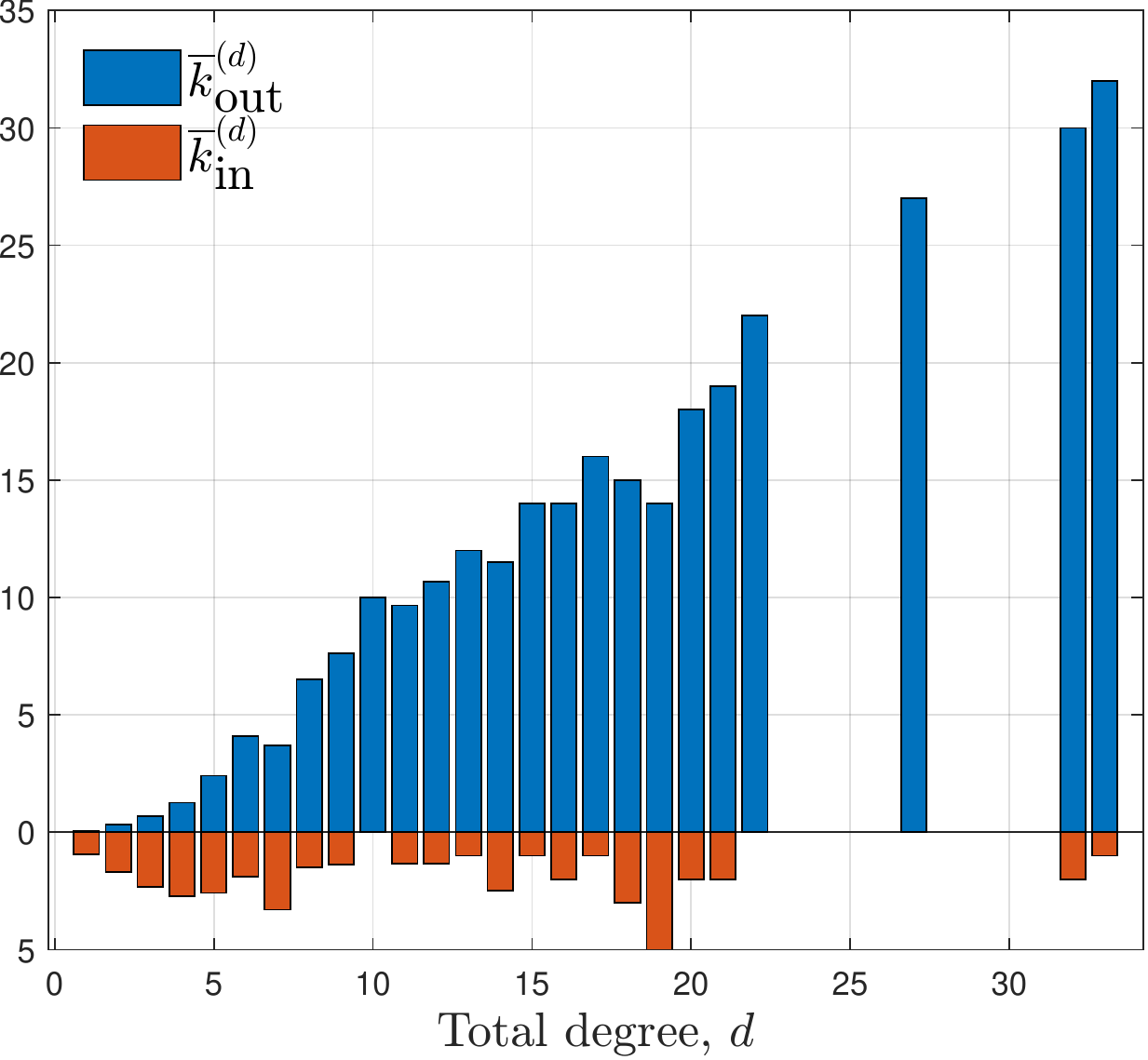}
	\caption{Decomposition $d=\bar{k}_\text{in}^{(d)}+\bar{k}_\text{out}^{(d)}$ of the total degree $d$ into the in- and out- components $\bar{k}_\text{in}^{(d)}$ and $\bar{k}_\text{out}^{(d)}$.}
	\label{fig:aveIN_vs_aveOUT}
\end{figure}

It is instructive to see how exactly the in- and out-degrees contribute to the total degrees of nodes. Let $\mathcal{G}_d\subset\mathcal{G}$ be the subset of nodes with total degree $d$ and $n_d$ be the number of nodes in $\mathcal{G}_d$,
\begin{equation}
\mathcal{G}_d=\{i~:~ k(i)=d\}, \hspace{5mm} n_d=|\mathcal{G}_d|.
\end{equation} 
Then for any node in $\mathcal{G}_d$ the sum of its in- and out-degrees is exactly $d$,
\begin{equation}\label{eq:kin+kout=d}
 k_\text{in}(i)+k_\text{out}(i)=d, \hspace{5mm}\forall i\in\mathcal{G}_d.
\end{equation}
Let $\bar{k}_\text{in}^{(d)}$ and $\bar{k}_\text{out}^{(d)}$ be the average in- and out-degrees of nodes in $\mathcal{G}_d$,
\begin{equation}
\bar{k}_\text{in}^{(d)}=\frac{1}{n_d}\sum_{i\in\mathcal{G}_d}k_\text{in}(i)\hspace{3mm}\mbox{and}\hspace{3mm}
\bar{k}_\text{out}^{(d)}=\frac{1}{n_d}\sum_{i\in\mathcal{G}_d}k_\text{out}(i).
\end{equation}
Then, by averaging (\ref{eq:kin+kout=d}) over all nodes $i\in\mathcal{G}_d$, we have
\begin{equation}\label{eq:kinAVE+koutAVE=d}
\bar{k}_\text{in}^{(d)}+\bar{k}_\text{out}^{(d)}=d,
\end{equation}
for any total degree $d$. Figure~\ref{fig:aveIN_vs_aveOUT} shows the decomposition of the total degree $d$ into the in- and out- components $\bar{k}_\text{in}^{(d)}$ and $\bar{k}_\text{out}^{(d)}$.  

In all subsets $\mathcal{G}_d\subset\mathcal{G}$ for $d>5$, $\bar{k}_\text{out}^{(d)}$ strongly dominates $\bar{k}_\text{in}^{(d)}$ and contributes the most to the total degree $d$. The in-degree component can be viewed as a ``noise'' added to the out-degree component.

Tables \ref{tab:top_in}, \ref{tab:top_out}, and \ref{tab:top_total} list the top courses with respect to the in-, out-, and total degrees.
\begin{table}[h]
	\centering
	\begin{tabular}{c|c|c|c}
			\multicolumn{4}{c}{\textsc{\textcolor{black}{In-Degree Top 9}}} \\
		\hline
		& \textsc{{Course}} & \textsc{{Title}} & \textsc{{In-Deg}} \\
		\hline
		{1}& CMS 139 & Algorithm Analysis & 7\\
		\hline
		{2}& Ge 270 & Continental Tectonics & 7\\
		\hline
		{3}& ACM 106 & Computational Math & 5\\
		\hline
		{4}& ChE 111 & Sustainability & 5\\
		\hline
		{5}& Ch 21 & Physical Chem & 5\\
		\hline
		{6}& Ch 25 & Biophysical Chem & 5\\
		\hline
		{7}& CMS 144 & Networks & 5\\
		\hline
		{8}& ME 50 & Modelling in Mech. Eng. & 5\\
		\hline
		{9}& Ph 6 & Physics Lab & 5
	\end{tabular}
	\caption{The top 9 courses with the largest in-degree. Cutoff was set to in-degree 5.}
	\label{tab:top_in}
\end{table}
\begin{table}[h]
	\centering
	\begin{tabular}{c|c|c|c}
		\multicolumn{4}{c}{\textsc{\textcolor{black}{Out-Degree Top 12}}} \\
		\hline
		& \textsc{{Course}} & \textsc{{Title}} & \textsc{{Out-Deg}} \\
		\hline
		{1}& Ma 2 & Differential Equations & 32\\
		\hline
		{2}& ACM 95/100 & Methods of Applied Math & 30\\
		\hline
		{3}& Ma 1 & Freshman Math & 27\\
		\hline
		{4}& Bi 8 & Molecular Bio & 22\\
		\hline
		{5}& Ma 3 & Intro Probability & 19\\
		\hline
		{6}& Ph 2 & Sophomore Physics & 19\\
		\hline
		{7}& Ph 125 & Quantum Mechanics & 17\\
		\hline
		{8}& Ch 41 & Organic Chem & 16\\
		\hline
		{9}& Ph 1 & Freshman Physics & 16\\
		\hline
		{10}& ACM 116 & Probability Models & 15\\
		\hline
		{11}& Ch 1 & Freshman Chem & 15\\
		\hline
		{12}& CS 1 & Intro Programming & 15
	\end{tabular}
	\caption{The top 12 courses with the largest out-degree. Cut-off was set to out-degree 15.}
	\label{tab:top_out}
\end{table}
\begin{table}[h]
	\centering
	\begin{tabular}{c|c|c|c}
		\multicolumn{4}{c}{\textsc{\textcolor{black}{Total Degree Top 7}}} \\
		\hline
		& \textsc{{Course}} & \textsc{{Title}} & \textsc{{Tot-Deg}} \\
		\hline
		{1}& Ma 2 & Differential Equations & 33\\
		\hline
		{2}& ACM 95/100 & Methods of Applied Math & 32\\
		\hline
		{3}& Ma 1 & Freshman Math & 27\\
		\hline
		{4}& Bi 8 & Molecular Bio & 22\\
		\hline
		{5}& Ph 2 & Sophomore Physics & 21\\
		\hline
		{6}& Ma 3 & Intro Probability & 20\\
		\hline
		{7}& Ph 125 & Quantum Mechanics & 20
	\end{tabular}
	\caption{The top 7 courses with the largest total degree. Cut-off was set to total degree 20.}
	\label{tab:top_total}
\end{table}

As expected, the top in-degree nodes in Table~\ref{tab:top_in} include some of the most specialized courses offered at Caltech. These courses are often at the graduate level and require a considerable amount of previous coursework in order to be taken. The top out-degree nodes in Table~\ref{tab:top_out} correspond to some of the most fundamental courses in mathematics, physics, biology, and chemistry that serve as prerequisites for many other courses. Note that some of these courses, such as Ma~1 and Ph~1, are introductory and elementary, but some, such as ACM~95/100, ACM~116, and Ph~125 are relatively advanced. These advanced courses serve as the basis for even more advanced courses, often at the graduate level. Table~\ref{tab:top_total} follows the order of Table~\ref{tab:top_out} closely. This confirms the dominance of the out-degree over the in-degree established in Fig.~\ref{fig:aveIN_vs_aveOUT}. 

Out of the three types of node degree, the \textit{out-degree} is the most suitable measure of course importance. Any node $i$ in a CPN fully controls its in-degree since the instructor who teaches course $i$ can choose which and how many other courses should be listed as prerequisites for $i$. Therefore, at least in principle, any node can make its in-degree as large as the entire network size. The in-degree measures the specialization of a course quantified by that \textit{single} course, but not its importance. The importance of a node should be determined by \textit{other} nodes, not by the node itself. This is exactly what happens for the out-degree since its value is determined by other courses: the out-degree of node $i$ is the number of other courses that decide to require $i$ as a prerequisite and establish links from $i$ to them. A node can ``control'' its out-degree only by being important for other nodes.

\subsection{PageRank Centrality}
\label{subsec:PageRankCentrality}

\begin{figure}[h]
	\centering
	\includegraphics[width = \linewidth]{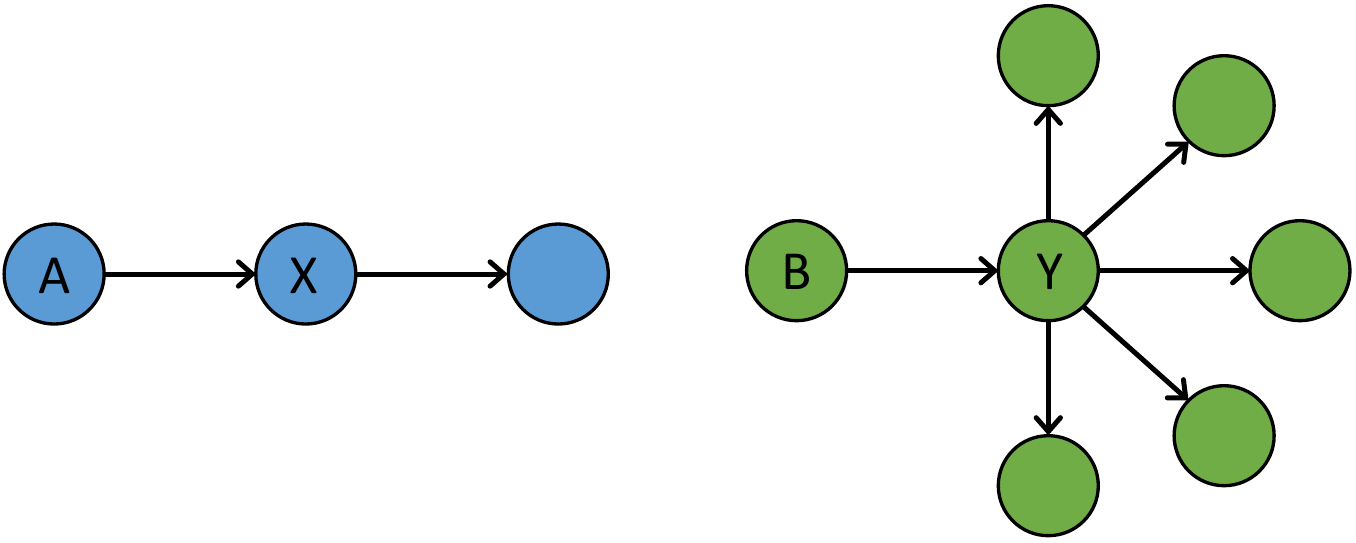}
	\caption{Intuitively, course B is more important than course A, but the out-degree does not distinguish  these two courses.}
	\label{fig:AandB}
\end{figure}

The out-degree is a simple and easily interpretable measure of course importance, but it has limitations and drawbacks. The out-degree of course $i$ gives it one ``importance point'' for every other course $j$ that lists $i$ as its prerequisite, but it does not take into account whether $j$ itself is an important course or not. As an example, consider two courses, A and B, in Fig.~\ref{fig:AandB}. 
The out-degree does not distinguish between these two courses and calls them equally important since $k_\text{out}(\text{A})=k_\text{out}(\text{B})=1$. Intuitively, however, course B is more important than A, since B is a prerequisite for Y, and Y has a larger out-degree than X (i.e., gives students more options for future coursework). Hence, the importance of a course $i$ is not only about how many courses use $i$ as a prerequisite, but also about how important the postrequisites of $i$ are. 

This idea is implemented in the \textit{PageRank} centrality measure~\cite{PageBrin1998}, which is a key ingredient of the Google search engine~\cite{BrinPage1998}. The PageRank centrality $\pi(i)$ of node $i$ is the sum of two terms. One term is a small amount $\phi$ of ``free'' importance that all nodes get regardless of their positions in the network. The other term is proportional to the sum of the PageRank centralities of $i$'s postrequisites, normalized by their in-degrees, so that 
\begin{equation}\label{eq:PageRankDef}
\pi(i)=\alpha \sum_{j=1}^n A_{ij}\frac{\pi(j)}{k_{\text{in}}(j)} + \phi.
\end{equation}
A course $j$ contributes to $\pi(i)$ if and only if $A_{ij}=1$, \ie $i$ is a prerequisite for $j$ (equivalently, $j$ is a postrequisite for $i$). Dividing $\pi(j)$ by $k_{\text{in}}(j)$ allows to prevent $j$ from being too influential by increasing the number of its prerequisites and, thus, contributing to $\pi(i)$ for many $i$. Any course $j$ that adopts this strategy of requiring many prerequisites will indeed contribute to many PageRank centralities $\pi(i)$, but its contribution $\pi(j)/k_{\text{in}}(j)$ will be small, since its in-degree  $k_{\text{in}}(j)$ will be large. The damping factor $\alpha\in(0,1)$ controls the relative contributions to $\pi(i)$ from the network structure and ``free centrality.'' If $\alpha=0$, then $\pi(i)=\phi$ for all nodes, which corresponds to the ultimately egalitarian case, where all nodes are equally important and network structure plays no role in determining node importance. As $\alpha$ increases, the influence of the network structure becomes stronger. The value of damping factor is usually set to $\alpha=0.85$, which is found to give good results~\cite{BrinPage1998}. The constant term $\phi$ ensures that any course $i$ with zero out-degree, \ie a ``dead end'' with  $A_{ij}=0$ for all $j$, still gets a non-zero centrality $\pi(i)=\phi\neq0$. This allows dead ends to contribute to PageRank centralities of their prerequisites. 

It can be shown (\eg \cite{Newman2018}, section 7.1.4)  that system (\ref{eq:PageRankDef}) of $n$ equations has a unique solution, and the vector $\pi=(\pi(1),\ldots,\pi(n))^\top$ of PageRank centralities is given by
\begin{equation}
\pi=\phi\left(I-\alpha AK_{\text{in}}^{-1} \right)^{-1}\mathbf{1},
\end{equation}
where $I$ is the $n\times n$ identity matrix, $\mathbf{1}=(1,\ldots,1)^\top$ is the $n\times 1$ vector of ones, and $K_{\text{in}}$ is the diagonal matrix with diagonal elements $K_{ii}=\max\{k_{\text{in}}(i),1\}$. A non-zero value of $\phi$ guarantees that $\pi\neq0$, but the exact value of $\phi$ does not affect the ranking based on PageRank centralities. This observation allows to conveniently set $\phi=1$, and get
\begin{equation}\label{eq:PageRank}
\pi=\left(I-\alpha AK_{\text{in}}^{-1} \right)^{-1}\mathbf{1}.
\end{equation}

It is worth mentioning that in the original definition of the PageRank centrality (\ref{eq:PageRankDef}), $A_{ji}$ and $k_{\text{out}}(j)$ are used instead of $A_{ij}$ and $k_{\text{in}}(j)$, respectively. As a result, in (\ref{eq:PageRank}), $A^\top$ and $K_{\text{out}}$ appear instead of $A$ and $K_{\text{in}}$. The reason for that is that the original version of PageRank was developed for the World Wide Web (WWW), where nodes (web pages) can control their outgoing links (hyperlinks to other web pages) but cannot control their incoming links (hyperlinks from other web pages), and where the importance of nodes is derived from incoming links. For CPNs the situation is exactly the opposite, and the roles of incoming and outgoing links, compared to the WWW, are switched. Therefore, (\ref{eq:PageRankDef}) and (\ref{eq:PageRank}) represent PageRank adapted to CPNs. The PageRank centralities of CPN nodes can be computed by using the standard PageRank method and applying it to $A^\top$, the transpose on CPN's adjacency matrix $A$. In this paper, all centrality measures are computed in \textit{NetworkX}, a Python package for analysis of  complex networks~\cite{NetworkX}. 
 
\begin{figure}[h]
	\centering
	\includegraphics[width = \linewidth]{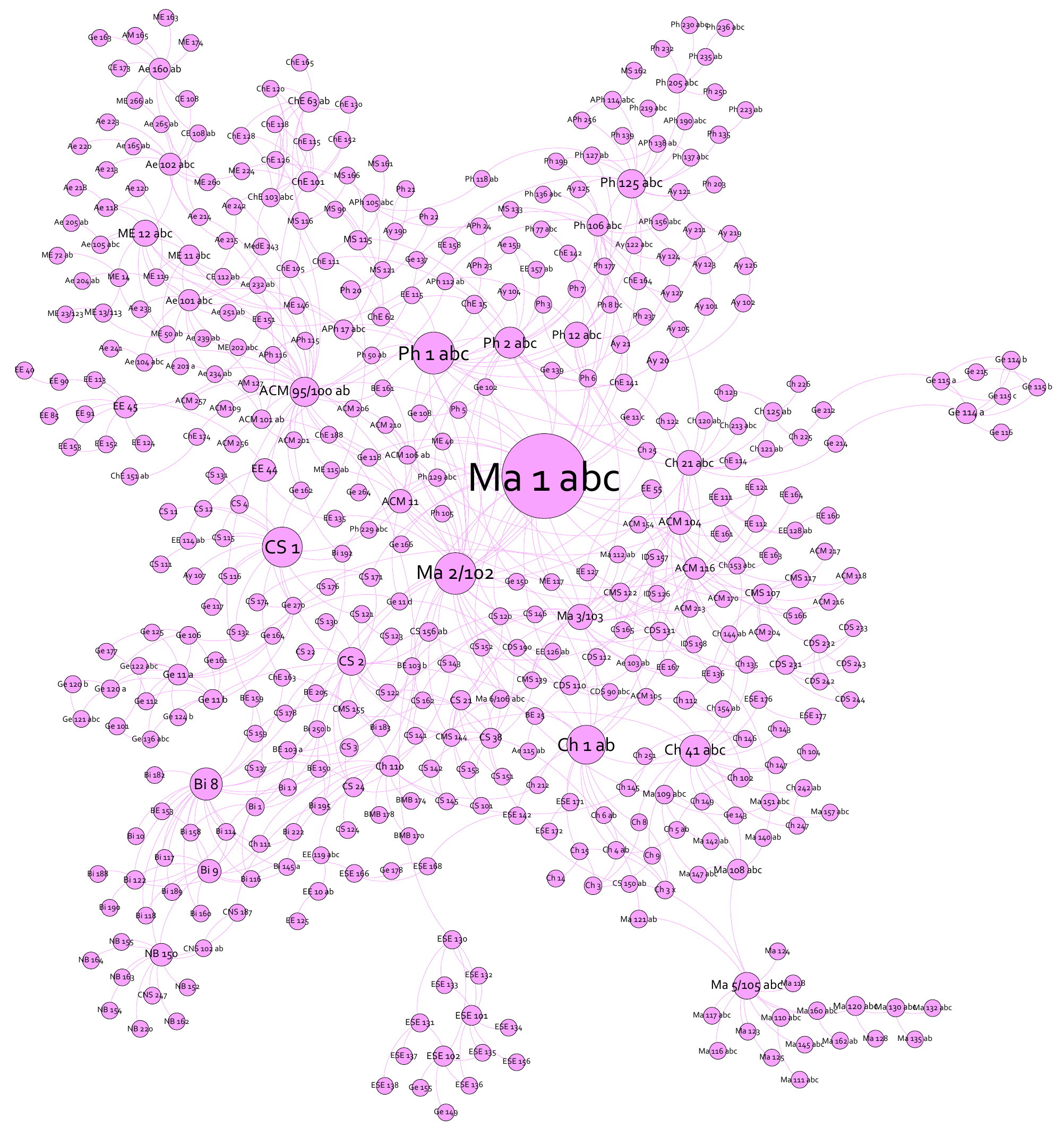}
	\caption{The LCC $\mathcal{G}$ of the Caltech CPN visualized with node size proportional to node's PageRank centrality: the larger the node, the higher its PageRank. For clarity, a large scale  visualization is shown in Fig.~\ref{fig:LCC_pr_full} in the Appendix.}
	\label{fig:LCC_pr}
\end{figure}

Figure~\ref{fig:LCC_pr} shows the largest connected component $\mathcal{G}$ of Caltech's CPN, where the size of node $i$ is proportional to its PageRank centrality $\pi(i)$. Table~\ref{tab:pagerank} lists the top 12 courses with respect to PageRank.
\begin{table}[h]
	\centering
	\begin{tabular}{c|c|c}
		\multicolumn{3}{c}{\textsc{\textcolor{black}{PageRank Top 12}}} \\
		\hline
		& \textsc{\textcolor{black}{Course}}  & \textsc{\textcolor{black}{Title}}\\
		\hline
		\textcolor{black}{1} & Ma 1 & Freshman Math\\
		\hline
		\textcolor{black}{2} & Ma 2 & Differential Equations\\
		\hline
		\textcolor{black}{3} & Ph 1 & Freshman Physics\\
		\hline
		\textcolor{black}{4} & CS 1 & Intro Programming\\
		\hline
		\textcolor{black}{5} & Ch 1 & Freshman Chem\\
		\hline
		\textcolor{black}{6} & Bi 8 & Molecular Bio\\
		\hline
		\textcolor{black}{7} & Ch 41 & Organic Chem\\
		\hline
		\textcolor{black}{8} & Ph 2 & Sophomore Physics\\
		\hline
		\textcolor{black}{9} & ACM 95/100 & Methods of Applied Math\\
		\hline
		\textcolor{black}{10} & Ph 125 & Quantum Mechanics\\
		\hline
		\textcolor{black}{11} & CS 2 & Data Structures \& Algorithms\\
		\hline
		\textcolor{black}{12} & Ma 5 & Intro Abstract Algebra
	\end{tabular}
	\caption{The top 12 courses with the largest PageRank.}
	\label{tab:pagerank}
\end{table}

\begin{figure}[h]
	\centering
	\includegraphics[width = \linewidth]{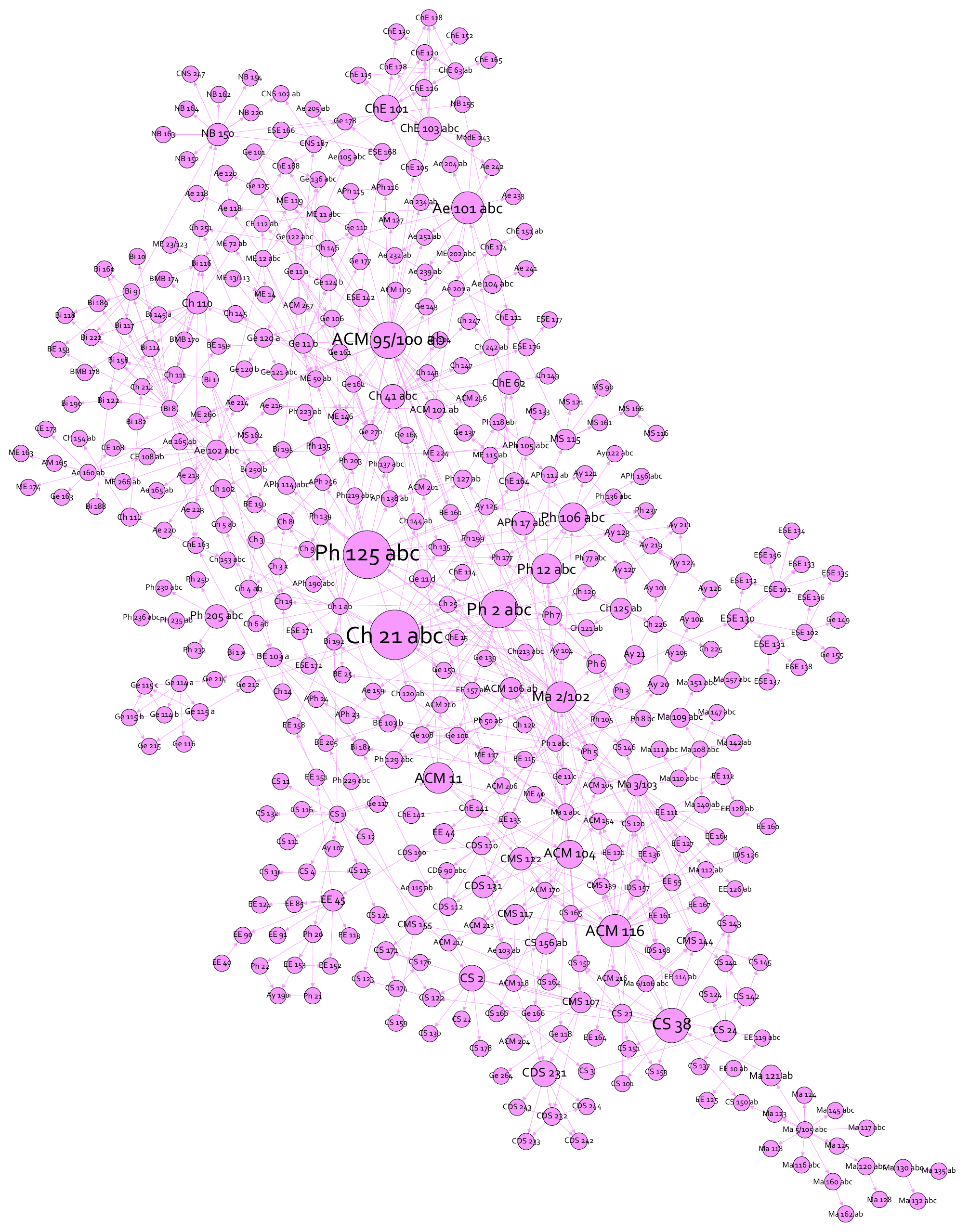}
	\caption{The LCC $\mathcal{G}$ of the Caltech CPN visualized with node size proportional to node's betweenness centrality: the larger the node, the higher its betweenness.}
	\label{fig:LCC_bet_full}
\end{figure}

It is interesting to compare the top 12 out-degree courses in Table~\ref{tab:top_out} with the top 12 PageRank courses in Table~\ref{tab:pagerank}. The first observation is that 10 out of 12 courses are present in both tables. This suggests that, similar to the out-degree, the PageRank centrality measures how \textit{fundamental} a course is. However, among those 10 common courses, PageRank tends to rank higher courses that are more \textit{introductory}. For example, PageRank prefers Ma~1 to Ma~2, Ph~1 to Ph~2, Ch~1 to Ch~41, and CS 1 to CS 2, while out-degree prefers Ma~2 to Ma~1, Ph~2 to Ph~1, Ch~41 to Ch~1, while CS 2 does not appear in the top 12 with respect to out-degree. The second course that made it to the top 12 according to PageRank but not according to out-degree is Ma~5, ``Introduction to Abstract Algebra.'' The out-degree of Ma~5 is 11, and 3 out of its postrequisites have positive out-degrees (see the bottom right tail of a network in Fig.~\ref{fig:LCC_bet_full}). This is also a case where PageRank can highlight certain characteristics of a university. Caltech's Mathematics department is very algebra oriented, with a lot of research carried out in algebraic fields and a lot of courses offered in advanced algebra topics. Due to this, Ma 5 is listed as a prerequisite for all such courses, thus boosting its place in the PageRank rankings.

PageRank's preference towards fundamental introductory courses can be explained as follows.  A course has a high PageRank centrality if it is either fundamental (has large out-degree) or less fundamental but serves as a prerequisite for fundamental courses. Consider now two courses in a CPN, A and B, which are approximately equally fundamental, $k_{\text{out}}(\text{A})\approx k_{\text{out}}(\text{B})$. If A is more introductory than B, then directed paths in the CPN starting from A are expected to be longer than directed paths starting from B. As a result, the postrequisites of A are likely to have higher PageRank scores than the postrequisites of B and, therefore, A is likely to have a higher PageRank than B, $\pi(\text{A})>\pi(\text{B})$.

Overall, in CPNs, the PageRank centrality measures how \textit{fundamental} a course is and favors more \textit{introductory} courses.

\subsection{Betweenness Centrality}
\label{subsec:BetweennessCentrality}

\begin{figure}[h]
	\centering
	\includegraphics[width = \linewidth]{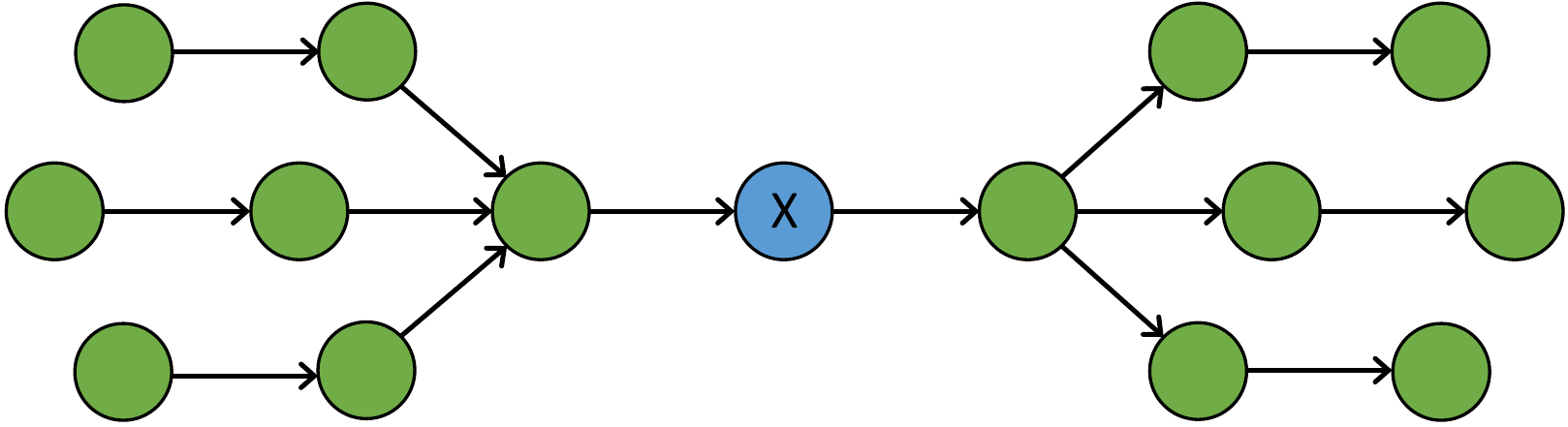}
	\caption{Important course X with small degree and PageRank.}
	\label{fig:courseX}
\end{figure}

A course-prerequisite network represents the flow of knowledge between different courses in a university curriculum, where knowledge ``flows'' along directed paths in the CPN from less to more advanced courses. A course with a small degree and low PageRank can still be important and influential if it occupies a dominant position with respect to the paths in the CPN. As an example, consider a course X in Fig.~\ref{fig:courseX}.  
Although this course has small in- and out-degrees and is poorly connected to the rest of the CPN, it serves as a \textit{bridge} between two subsets of courses. Any path from the left subset to the right must go through course X. This gives X control over the flow of knowledge between other courses and makes it important. 

This notion of importance is formalized in the \textit{betweenness} centrality~\cite{Anthonise1971,Freeman1977}. Mathematically, the  betweenness centrality $\beta(i)$ of node $i$ in a CPN is defined as follows. Let $\sigma(s,t)$ be the total number of shortest paths from node $s$ to node $t$, and $\sigma(s,t\hspace{0.3mm}|\hspace{0.3mm}i)$ be the number of shortest paths from $s$ to $t$ that pass through node $i$. All paths are understood as directed paths that go along directed links. Note that if $\sigma(s,t)>0$, \ie there exists a  path from $s$ to $t$, then $\sigma(t,s)=0$, since a CPN does not have loops. The betweenness centrality of node $i$ is
\begin{equation}\label{eq:betDef}
\beta(i)=\sum_{s\neq i, t\neq i}\frac{\sigma(s,t\hspace{0.3mm}|\hspace{0.3mm}i)}{\sigma(s,t)},
\end{equation}
where the sum is over all nodes $s,t$ in the CPN that are different from $i$, with a convention that if $\sigma(s,t)=\sigma(s,t\hspace{0.3mm}|\hspace{0.3mm}i)=0$, then $\sigma(s,t\hspace{0.3mm}|\hspace{0.3mm}i)/\sigma(s,t)=0$. For example, the betweenness of course X in Fig.~\ref{fig:courseX} is $\beta(\text{X})=49$. For a connected pair of nodes $(s,t)$, \ie for nodes with $\sigma(s,t)\neq0$, the ratio $\sigma(s,t\hspace{0.3mm}|\hspace{0.3mm}i)/\sigma(s,t)$ can be interpreted as the probability that a shortest path from $s$ to $t$, chosen uniformly at random from all shortest paths from $s$ to $t$, passes through $i$. 

The betweenness centrality quantifies the degree to which a node is located \textit{between} other nodes. If a node has zero in-degree (``source'') or zero out-degree (``sink'' or ``dead end''), then, by definition (\ref{eq:betDef}), its betweenness is zero, 
\begin{equation}
k_{\text{in}}(i)=0 \hspace{3mm}\textrm{or}\hspace{3mm} k_{\text{out}}(i)=0 \hspace{3mm}\implies\hspace{3mm} \beta(i)=0.
\end{equation}

For course-prerequisite networks, betweenness has the following interesting interpretation. Courses with high betweenness are \textit{intermediate-level} courses that serve as \textit{critical bridges} between less and more advanced courses. Removing them from a CPN (by halting their teaching) would greatly disrupt the flow of knowledge in the CPN. 

\begin{figure}[h]
	\centering
	\includegraphics[width = \linewidth]{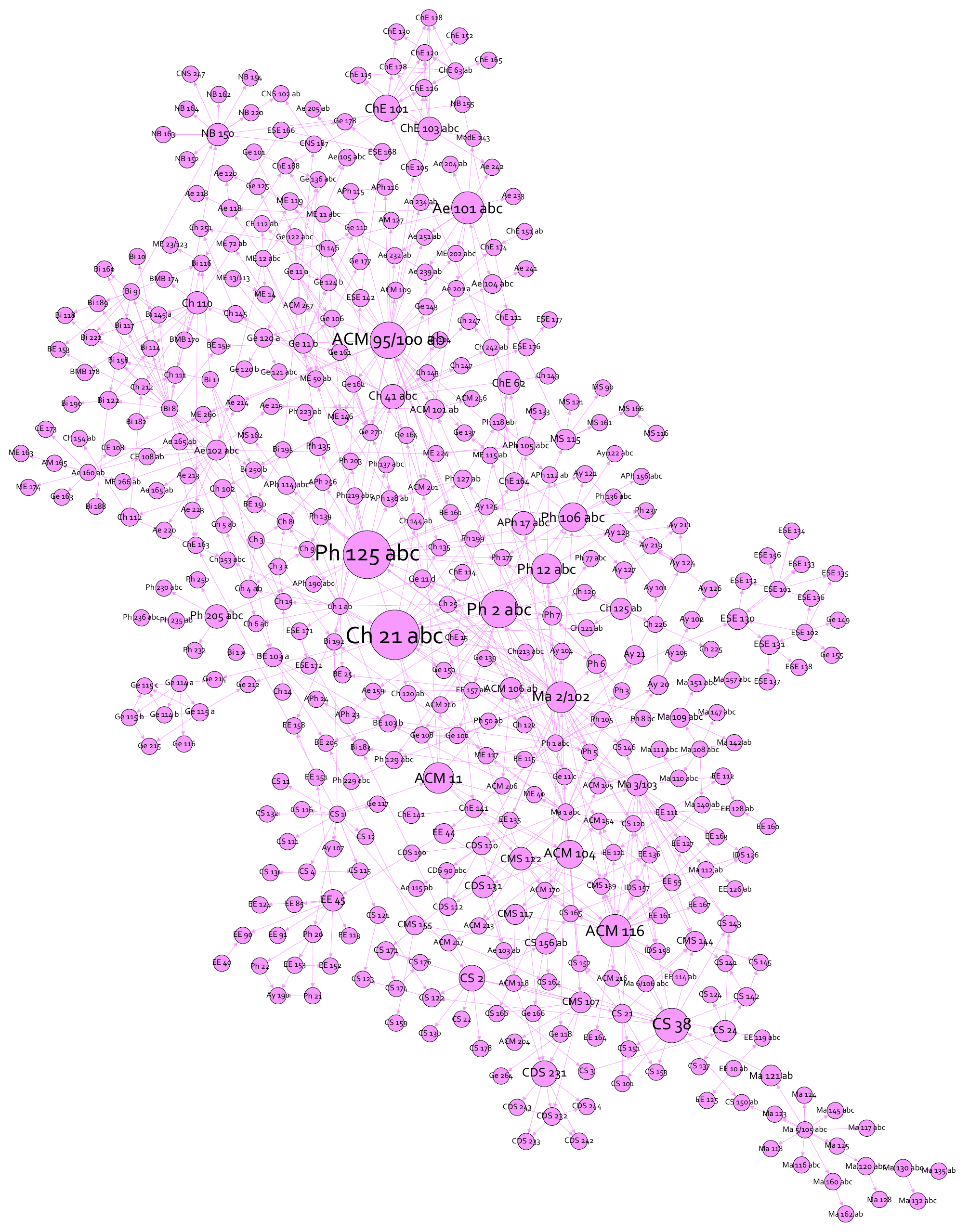}
	\caption{The LCC $\mathcal{G}$ of the Caltech CPN visualized with node size proportional to node's betweenness centrality: the larger the node, the higher its betweenness. For visual clarity, a large scale  visualization is shown in Fig.~\ref{fig:LCC_bet_full} in the Appendix.}
	\label{fig:LCC_bet}
\end{figure}

Figure~\ref{fig:LCC_bet} shows the LCC $\mathcal{G}$ of Caltech's CPN, where the size of node $i$ is proportional to its betweenness centrality $\beta(i)$. Table~\ref{tab:Betweenness} lists the top 12 courses with respect to the betweenness.
\begin{table}[h]
	\centering
	\begin{tabular}{c|c|c}
		\multicolumn{3}{c}{\textsc{\textcolor{black}{Betweenness Top 12}}} \\
		\hline
		& \textsc{\textcolor{black}{Course}}  & \textsc{\textcolor{black}{Title}}\\
		\hline
		\textcolor{black}{1} & Ch 21 & Physical Chem\\
		\hline
		\textcolor{black}{2} & Ph 125 & Quantum Mechanics\\
		\hline
		\textcolor{black}{3} & Ph 2 & Sophomore Physics\\
		\hline
		\textcolor{black}{4} & ACM 95/100 & Methods of Applied Math\\
		\hline
		\textcolor{black}{5} & CS 38 & Algorithms\\
		\hline
		\textcolor{black}{6} & ACM 116 & Probability Models\\
		\hline
		\textcolor{black}{7} & Ma 2 & Differential Equations\\
		\hline
		\textcolor{black}{8} & ACM 11 & Intro Computational Science\\
		\hline
		\textcolor{black}{9} & Ph 12 & Analytical Sophomore Physics\\
		\hline
		\textcolor{black}{10} & Ph 106 & Topics in Classical Physics\\
		\hline
		\textcolor{black}{11} & ACM 104 & Applied Linear Algebra\\
		\hline
		\textcolor{black}{12} & Ae 101 & Fluid Mechanics
	\end{tabular}
	\caption{The top 12 nodes with the largest betweenness.}
	\label{tab:Betweenness}
\end{table}

As expected, all high betweenness courses are important intermediate-level courses that equip students with fundamental (but not elementary) concepts and tools in mathematics, physics, chemistry, and computer science, in preparation for more advanced courses. Interestingly, many high betweenness courses are \textit{interdisciplinary}. For example, Ch~21 combines physics and chemistry, ACM~95/100 teaches methods of applied mathematics (complex analysis, ordinary differential equations, partial differential equations) for the physical sciences, and ACM~116 is a course on probability models for science and engineering students. This suggests that courses with high betweenness do not only serve as bridges between less and more advanced courses, but are also likely to be interdisciplinary and combine different traditional academic disciplines. 

Among the top 12 PageRank courses in Table~\ref{tab:pagerank} and top 12 betweenness courses in Table~\ref{tab:Betweenness}, there are 4 courses that appear in both tables: Ph~125, Ph~2, ACM~95/100, and Ma~2. These are important, fundamental courses at the intermediate level. The presence of courses with high values of both centrality measures suggests that PageRank and betweenness are positively correlated. Indeed, for the Caltech CPN the Pearson correlation coefficient between PageRank and betweenness is
\begin{equation}
    \rho_{\pi,\beta}=0.34.
    \label{eq:rho_pr_bet}
\end{equation}  

\begin{figure}[t]
	\centering
	\includegraphics[width = \linewidth]{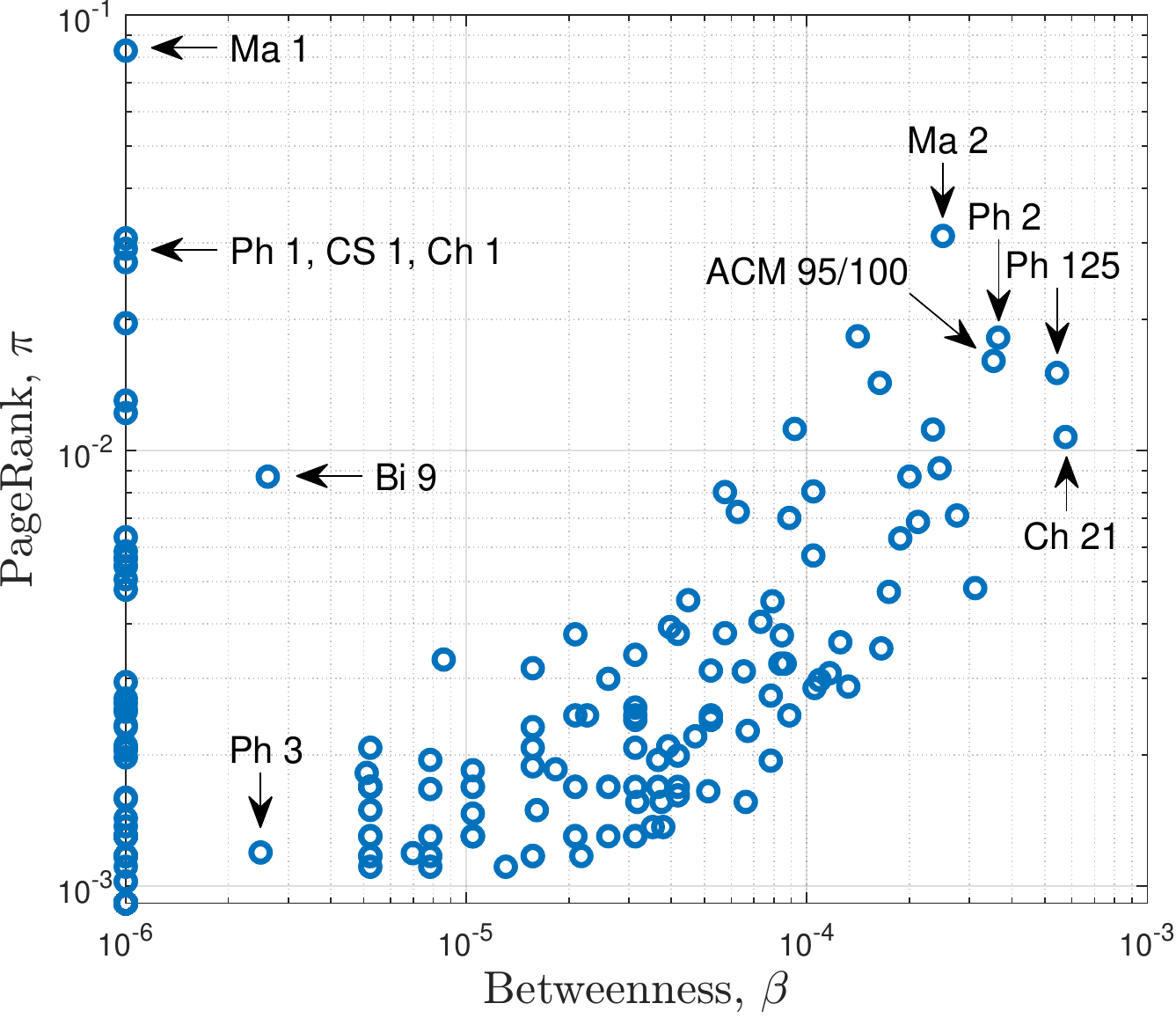}
	\caption{Scatter plot of PageRank versus betweenness for the LCC $\mathcal{G}$ of the Caltech CPN. Courses that lie on the vertical line at the far-left side of the plot have zero betweenness. }
	\label{fig:scatterplot_pr_bet}
\end{figure}

Figure~\ref{fig:scatterplot_pr_bet} shows the scatter plot of the two centrality measures. There is a noticeable positive correlation between PageRank and betweenness, especially for courses with large values of PageRank or betweenness. This means that courses which are important with respect to one centrality measure tend to be also important with respect to the other. 
\begin{figure}[h]
	\centering
	\includegraphics[width = \linewidth]{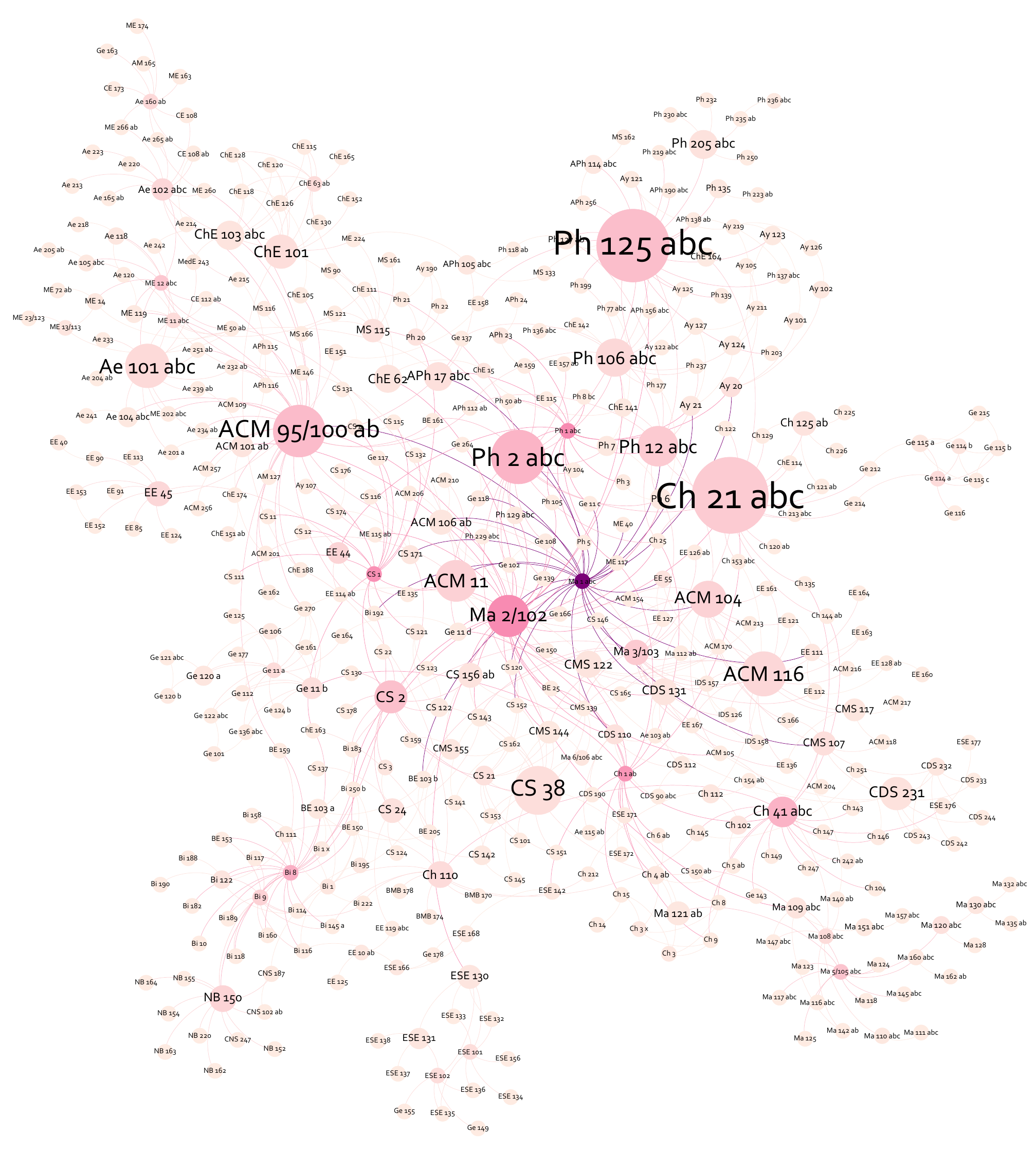}
	\caption{The LCC $\mathcal{G}$ of the Caltech CPN visualized with node size proportional to node's betweenness and color intensity proportional to node's PageRank. The larger and darker the node, the higher its betweenness and PageRank, the more important it is. Links are directed and start at the node of the same color. For clarity, a large scale  visualization is shown in Fig.~\ref{fig:LCC_pr_bet_full} in the Appendix.}
	\label{fig:LCC_pr_bet}
\end{figure}

To better visualize the interplay between PageRank and betweenness, Fig.~\ref{fig:LCC_pr_bet} shows the LLC $\mathcal{G}$ of the Caltech CPN, where node size is proportional to node's betweenness and the color intensity is proportional to node's PageRank, so that large dark nodes are important with respect to both centrality measures. 
As expected in view of (\ref{eq:rho_pr_bet}), on average, larger nodes tends to be darker. One notable exception is the small dark node in the middle of Fig.~\ref{fig:LCC_pr_bet}, which corresponds to Ma~1, Freshman Math. It has the highest PageRank in $\mathcal{G}$, but its betweenness is zero since its in-degree is zero.

\section{Topological Stratification}
\label{sec:TopologicalOrdering}

A course-prerequisite network is a directed network with a special property: it does not have cycles (or loops). A cycle is a closed directed path, \ie a path that starts and ends at the same node and goes along links only in their forward direction. Absence of cycles means that if there is a directed path from node $i$ to node $j$, then there is no a directed path from $j$ to $i$. A directed network without cycles is called a \textit{directed acyclic graph} (DAG). %So, any course-prerequisite network is a DAG.

Directed acyclic graphs constitute an important class of directed networks and have many scientific, statistical, and computational applications in various fields. Examples of complex systems which are modeled as DAGs include citation networks~\cite{Radicchi2009}, where nodes represent academic papers and links represent citations between them, and food webs~\cite{Dunne2002}, where nodes represent species in an ecosystem and links represent predator-prey relationships. In the causal set approach to quantum gravity, DAGs are used to describe the discrete structure of spacetime~\cite{Bombelli1987}. Bayesian networks~\cite{Cowel1999} and feedforward neural networks~\cite{Goodfellow2016}, which are widely used in causal inference and machine learning, are other important examples of directed acyclic graphs.  

The underlying reason why any CPN is a DAG is that CPN nodes representing courses are \textit{partially time-ordered}: any course can be taken only after its prerequisites have been taken. Existence of a cycle $i\rightarrow j \rightarrow k\rightarrow\ldots\rightarrow i$ would mean that course $i$ is an indirect prerequisite for itself and, therefore, $i$ can be taken only after $i$ has been taken, which is absurd. The acyclic structure of a CPN is thus a direct corollary of the time-ordering. The time-ordering is partial since if nodes $i$ and $j$ are not connected by a directed path, \ie neither course serves as an indirect prerequisite for the other, then courses $i$ and $j$ can be taken in any order.

\begin{figure}[t]
	\centering
	\includegraphics[width = 0.9\linewidth]{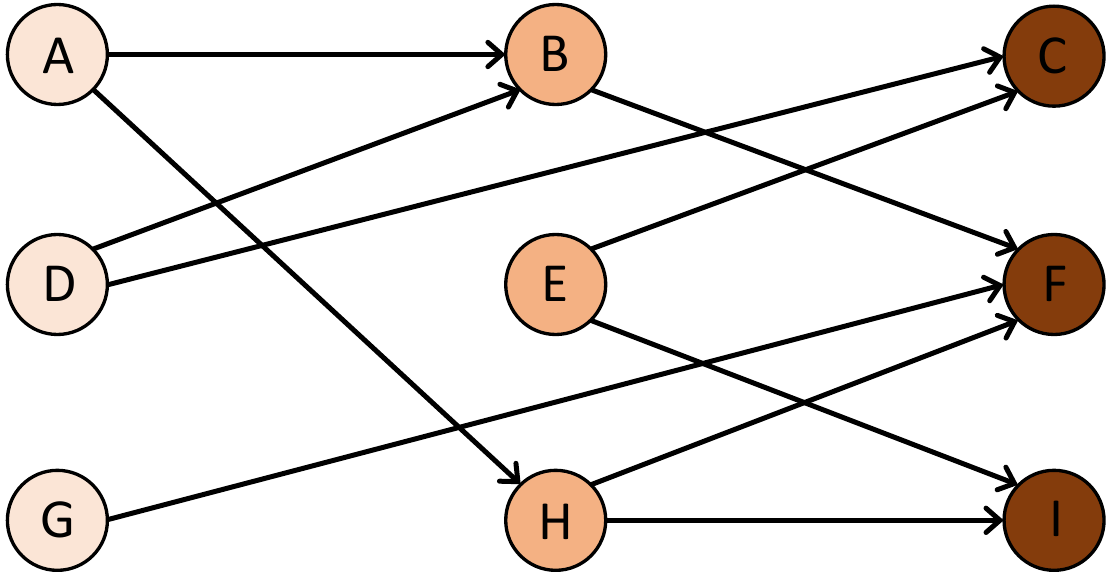}
	\caption{An example of a directed acyclic graph (DAG).}
	\label{fig:DAG}
\end{figure}

The acyclic structure of a DAG allows to topologically order its nodes. A \textit{topological ordering} of a directed network is an ordering of its nodes $i_1<i_2<\ldots<i_n$ such that for each link $i_k\rightarrow i_m$ from node $i_k$ to node $i_m$, we have $i_k<i_m$, \ie $i_k$ appears before $i_m$ in the ordering. To illustrate this definition, consider a DAG on Fig.~\ref{fig:DAG}.
This graph has several topological orderings. For example,
\begin{equation}\label{eq:toporder_example}
\begin{split}
\text{A $<$ D $<$ G $<$ B} &\text{ $<$ E $<$ H $<$ C $<$ F $<$ I}\\
&\text{and}\\
\text{D $<$ A $<$ B $<$ H} &\text{ $<$ G $<$ F $<$ E $<$ I $<$ C}
\end{split}
\end{equation}
are two of them. It can be shown that a directed network has a topological ordering if and only if it is a DAG, and the topological ordering is unique if and only if the network has a Hamiltonian path (a directed path that visits each node exactly once)~\cite{DAG}. Real-world CPNs do not have Hamiltonian paths, since otherwise would imply that for any pair of courses one is an indirect prerequisite for the other, which is not realistic. As a result, real-world CPNs allow multiple topological orderings. 

A topological ordering of a CPN is a traversal of the network in which each course is visited only after all its prerequisites have been visited. This linear ordering yields a \textit{ranking} of CPN nodes. The ranking induced by a topological ordering has, however, a serious drawback: it imposes a total order on partially ordered CPN nodes by arbitrary ordering \textit{topologically equivalent} nodes, \ie nodes that are not connected by a directed path (courses that are not indirect prerequisites for each other)~\cite{Zinoviev2018}. This drawback stems from the non-uniqueness of topological ordering. For example, the first topological ordering in (\ref{eq:toporder_example}) ranks D higher than A, and the second ordering ranks A higher than D, although A and D are topologically equivalent.

The arbitrariness of ranking can be remedied as follows. Instead of imposing a total order on CPN nodes, we can partition the nodes into disjoint subsets, or \textit{strata}, of topologically equivalent nodes. The last stratum $\mathcal{S}_T$ is a subset of nodes with zero out-degree, \ie the most advanced courses without any postrequisites (sinks or dead ends). The second to last stratum $\mathcal{S}_{T-1}$ is obtained by first removing $\mathcal{S}_T$ together with their incoming links and then finding nodes with zero out-degree in the remaining network. This process continues until all nodes are assign to their strata and the CPN is partitioned into $T$ strata, 
\begin{equation}\label{eq:topstrat}
S(\mathcal{G})=\{\mathcal{S}_1,\mathcal{S}_2,\ldots,\mathcal{S}_T\}.
\end{equation}
We refer to this ordered partition $S(\mathcal{G})$ as the \textit{topological stratification} of CPN $\mathcal{G}$. Each stratum $\mathcal{S}_t$ consists of topologically equivalent and unordered nodes, but the partition itself is ordered. For the network in Fig.~\ref{fig:DAG}, the topological stratification looks as follows:
\begin{equation}\label{eq:topstrat_example}
\begin{split}
\mathcal{S}_3&=\{\text{C, F, I}\},\\
\mathcal{S}_2&=\{\text{B, E, H, G}\},\\
\mathcal{S}_1&=\{\text{A, D}\},\\
S(\mathcal{G})&=\{\{\text{A, D}\},\{\text{B, E, H, G}\},\{\text{C, F, I}\}\}.
\end{split}
%\{\text{A, D, G}\}\sqcup\{\text{B, E, H}\}\sqcup\{\text{C, F, I}\}.
\end{equation}

The process of finding the topological stratification of a CPN resembles a slicing process: the nodes are ``sliced'', stratum-by-stratum, starting from the most advanced courses at the ``top'' of the network to the most elementary courses at the ``bottom''. There is also an alternative way $\widetilde{S}$ of partitioning CPN nodes into disjoint subsets of topologically equivalent nodes, which is, in some sense, the reverse of the process described above. Namely, the first subset is a set of nodes with zero in-degree. The second subset is obtained by removing the first subset of nodes together with their outgoing links and then finding nodes with zero in-degree in the remaining network, and so on. Note that, in general, the partition $\widetilde{S}(\mathcal{G})$ obtained by this method is different form the partition (\ref{eq:topstrat}) obtained via topological stratification. For example, for the network in Fig.~\ref{fig:DAG}, this alternative method leads to
\begin{equation}
\widetilde{S}(\mathcal{G})=\{\{\text{A, D, G, E}\},\{\text{B, H, C}\},\{\text{F, I}\}\},
\end{equation}
which is different from (\ref{eq:topstrat_example}). 

The relationship between the method of topological stratification $S$ and the alternative partitioning method $\widetilde{S}$ is straightforward. Let $\mathcal{G}'$ denote the reverse of ${\mathcal{G}}$, \ie a network obtained from ${\mathcal{G}}$ by reversing the directions of all its links. Furthermore, let $\overleftarrow{\hspace{5mm}}$ be the operation of reversing the order of a partition, \ie 
\begin{equation}
\overleftarrow{\{\mathcal{S}_1,\ldots,\mathcal{S}_T\}}=\{\mathcal{S}_T,\ldots,\mathcal{S}_1\},
\end{equation}
then 
\begin{equation}
\widetilde{S}(\mathcal{G})=
\overleftarrow{S(\mathcal{G}')}. 
\end{equation}

The reason why  the topological stratification of $\mathcal{G}$ is defined as $S(\mathcal{G})$ and not as $\widetilde{S}(\mathcal{G})$, that is, why we start from nodes with zero out-degree rather than nodes with zero in-degree is the following. While nodes with zero out-degree are indeed advanced courses (all have at least one direct prerequisite and most have many indirect prerequisites), nodes with zero in-degree are not necessarily elementary. For example, first-year graduate courses are quite advanced, but they can have zero in-degree in the CPN, since they may not have any formal prerequisites (only informal ones, \ie ``knowledge of quantum mechanics and probability theory at the undergraduate level is assumed''). As it follows from the subsequent discussion, combining those advanced zero in-degree courses with truly elementary courses (this is exactly what $\widetilde{S}$ does) is undesirable.

Topological stratification (\ref{eq:topstrat}) induces a \textit{hierarchical structure} on a CPN 
\begin{equation}\label{eq:hierarchicalstructure}
\mathcal{S}_1\prec\mathcal{S}_2\prec\ldots\prec\mathcal{S}_T,
\end{equation}
where relation $\mathcal{S}_{i}\prec\mathcal{S}_{i+1}$ means that any course in $\mathcal{S}_{i+1}$ can be taken after all courses in $\mathcal{S}_{i}$ have been taken. All courses in stratum $\mathcal{S}_i$ have approximately the same level of difficulty, and they are more advanced than courses in the previous stratum $\mathcal{S}_{i-1}$ and less advanced than courses in the next stratum $\mathcal{S}_{i+1}$. So, topological stratification of a CPN allows to view the hierarchy of the whole university curriculum with respect to the level of advancement.  

Topological stratification describes the ``meso-scale'' (intermediate-scale) organization of course-prerequisite networks or, more generally, directed acyclic graphs. Other types of meso-scale structures in complex networks include \textit{community structure}~\cite{GrivanNewman2002,Porter2009,Fortunato2010} that identifies densely connected groups of nodes (``communities'') that are sparsely connected between each other, \textit{core–periphery structure}~\cite{Holme2005,Csermely2013,Rombach2014} that divides a network into a densely connected core and a sparser periphery that is well connected to the core, and the $k$\textit{-core decomposition}~\cite{Alvarez2005} together with its refinement, the \textit{onion decomposition}~\cite{Dufresne2016}, that partition a network into a sequence of nested cores. A common feature of all these meso-scale structures is that they partition the network nodes into a collection of subsets with respect to a certain criterion. For the aforementioned structures, this criterion quantifies the density/sparsity of links between nodes or the ``coreness'' of the nodes. For the topological stratification, which is defined only for DAGs, this criterion is the topological equivalence of nodes, that is, the absence of directed paths between them. In the context of CPNs, the topological stratification yields subsets of courses of approximately the same level of difficulty that are not indirect prerequisites of each other.

Table~\ref{tab:Stratification} in the Appendix shows the topological stratification of the LLC $\mathcal{G}$ of Caltech's CPN obtained by using Python module~\cite{toposort}. The resulting stratification has $T=7$ strata, starting with the most elementary undergraduate courses and building up to the most advanced courses, often at the graduate level.
  
The topological stratification provides a hierarchical representation of the curriculum, which can be used by students for making more informed choices about what courses they can and should take to move up a level in the hierarchy. Moreover, the topological stratification unveils \textit{hidden prerequisites}, \ie courses that are not explicitly listed as official prerequisites for a given course, but would be beneficial if completed before taking the course. For example, ACM~104 (Applied Linear Algebra) is placed in the stratum below ACM~95/100 (Methods of Applied Math),   
\begin{equation}\label{eq:ACM104_ACM95100}
\text{ACM~104}\in\mathcal{S}_3\prec\mathcal{S}_4\ni\text{ACM~95/100}.
\end{equation}
The list of official prerequisites for ACM~95/100 does not contain ACM~104. However, relation (\ref{eq:ACM104_ACM95100}) suggests that ACM~104 should be taken before ACM~95/100. This makes sense because ACM 95/100 covers ordinary and partial differential equations, and being familiar with the fundamental concepts of linear algebra such as linear dependence and independence, span, basis, eigenvalues, eigenvectors, and matrix theory is extremely helpful. These topics are covered to some degree in $\text{Ma~1~abc}\in\mathcal{S}_1$, which is an official prerequisite for ACM~95/100, but in ACM~104 they are discussed at a deeper level. The first author, who took both courses, and the second author, who teaches both courses, confirm that ACM~95/100 is a more advanced course, and it would indeed be beneficial to take ACM~104 before taking ACM~95/100.

The hierarchical structure (\ref{eq:hierarchicalstructure}) induces a \textit{comprehensive schedule} in which students can take courses offered by their major department. Let $\mathcal{M}\subset\mathcal{G}$ be a subset of courses offered by a specific major, such as  ``ACM'' (Applied and Computational Mathematics) or ``IDS'' (Information and Data Sciences). Then (\ref{eq:hierarchicalstructure}) reduced on $\mathcal{M}$ gives
\begin{equation}\label{eq:hierarchicalstructureM}
(\mathcal{S}_1\cap\mathcal{M})\prec(\mathcal{S}_2\cap\mathcal{M})\prec\ldots\prec(\mathcal{S}_T\cap\mathcal{M}).
\end{equation}
This reduced hierarchy suggests that students interested in a deep and thorough  understanding of their major should first take courses from $\mathcal{S}_1\cap\mathcal{M}$, then courses from $\mathcal{S}_2\cap\mathcal{M}$, etc. Courses from $\mathcal{S}_t\cap\mathcal{M}$ can be taken in any available order and, perhaps, complemented with other relevant courses from $\mathcal{S}_t$. This schedule guarantees that all prerequisites from $\mathcal{M}$ are satisfied before taking a course that is higher in the hierarchy. Furthermore, as demonstrated above, the comprehensive schedule can recommend hidden prerequisites and provide students who adopt it with confidence that they are as prepared as possible for taking more advanced courses. 

Courses that belong to the same stratum $\mathcal{S}_t$ are topologically equivalent and generally offered by different university departments. They can be viewed as the \textit{building blocks} for all courses in $\mathcal{S}_{t'}$ for all $t'>t$, which are also offered by different departments. This creates an intriguing interdependence between departments, which is investigated in the next section.

\section{Interdependence Analysis}
\label{sec:InterdependenceAnalysis}

The term \textit{interdisciplinary} has become almost synonymous with contemporary research. New research areas constantly emerge through the interaction between two or more academic fields. For example, the recent development of computational, information, and data sciences has allowed for computational thinking and methods to be integrated with numerous traditional academic disciplines, giving rise to new, interdisciplinary fields. To successfully work in the emerging interdisciplinary areas, scientists need to have a broad skill-set and should be able to collaborate with colleagues from different departments. Since one of the main goals of modern universities is to prepare the new generation of scientists, we expect this trend to manifest itself in university curricula as well, reflecting the interdisciplinary nature of contemporary science, engineering, and technology. To quantify the degree to which this is the case, we turn to what we refer to as an \textit{interdependence analysis}.

Interdependence encapsulates whether, and to what extent, courses in one area of study have prerequisites from other areas. For example, a course on algorithms offered by the computer science department may have a discrete math prerequisite offered by the mathematics department. Interdependence also demonstrates the flow of knowledge from one area of study to another, and shows what ``external'' areas a student is expected to be familiar with in order to study a chosen academic discipline. To formalize this in the context of CPNs, we first need to identify the existing areas of study within the university. For the Caltech CPN, we consider 6 academic divisions (Table~\ref{tab:Divisions}) that offer courses in 25 areas of study, formalized in the corresponding degree options,  (Table~\ref{tab:Departments}). 
\begin{table}[h]
	\centering
	\begin{tabular}{c|c}
		\textbf{Division} & \textbf{Name}  \\
		\hline
		PMA &  Physics, Mathematics, Astronomy\\
		\hline
		EAS & Engineering \& Applied Science\\
		\hline
		CMS-div & Computing \& Mathematical Sciences\\
		\hline
		CHCHE & Chemistry \& Chemical Engineering\\
		\hline
		BBE & Biology \& Biological Engineering\\
		\hline
		GPS & Geology \& Planetary Sciences
	\end{tabular}
	\caption{Caltech academic divisions in the CPN.}
	\label{tab:Divisions}
\end{table}
\begin{table}[h]
    \centering
    \begin{tabular}{c|c}
        \textbf{Division} & \textbf{Areas of Study} \\
        \hline
        PMA & Ph, Ma, Ay\\
        \hline
        EAS & Ae, APh, MS, EE, ME, MedE, CE, AM\\
        \hline
        CMS-div & CMS, ACM, CS, IDS, CDS\\
        \hline
        CHCHE & Ch, ChE\\
        \hline
        BBE & BMB, BE, Bi, CNS, NB\\
        \hline
        GPS & ESE, Ge
    \end{tabular}
    \caption{Areas of study arranged by their respective division. The full names and descriptions of all listed areas are available in the 2021-22 Caltech Catalog~\cite{CaltechCatalog}.}
    \label{tab:Departments}
\end{table}

Two important remarks regarding the divisions listed in~Table~\ref{tab:Divisions} are in order. First, although CMS-div is officially a department in the EAS division, we consider it as a separate standalone division for the purposes of interdependence analysis. The reason is twofold: a) while other EAS departments focus on applied science and engineering, CMS-div revolves around computational and mathematical sciences, and b) EAS considered together with CMS-div would be too large, making the interdependence analysis at the division level meaningless.  It is also important to distinguish between CMS-div and CMS: CMS-div is a division and CMS is one of the areas of studies offered by CMS-div. Second, Caltech has an additional division, HSS (Humanities \& Social Sciences), which is excluded from the interdependence analysis because almost all its courses have no prerequisites and postrequisites, \ie they are isolated nodes in the Caltech CPN and are not in the LCC $\mathcal{G}$.

We define the interdependence between areas of study and between academic divisions as follows. Let $\mathcal{D}$ be the set of all divisions and $\mathcal{A}$ be the set of all areas of study within the university. For every area of study $a\in\mathcal{A}$ there exists a division $d\in\mathcal{D}$ that offers courses in that area, denoted $a\in d$. For example, $\textrm{Ma}\in\textrm{PMA}$ and $\textrm{Bi}\in\textrm{BBE}$. Let $\mathcal{C}_a\subset\mathcal{G}$ be the set of all courses in area $a\in\mathcal{A}$. In terms of the CPN nodes, 
\begin{equation}
\mathcal{C}_a=\{i\in\mathcal{G}:\hspace{1mm} a_i=a\},
\end{equation}
where $a_i$ is the area of node (course) $i$, \eg the areas of ACM~104 and IDS~157 are ACM and IDS, respectively. The number of courses in $\mathcal{C}_a$ is denoted by $|\mathcal{C}_a|$.

Furthermore, let $\mathcal{C}_a^+\subset\mathcal{G}$ be the \textit{multiset} of all postrequisites of $\mathcal{C}_a$. That is, $\mathcal{C}_a^+$ consists of all courses that are pointed-to by the courses in $\mathcal{C}_a$, and (this is the difference between a ``set'' and a ``multiset'') each course can occur more than once. As an example, consider the toy CPN in Fig.~\ref{fig:toyCPN}, and let the top row of courses represent the area of study $a=\{\text{A}, \text{B}, \text{C}\}$. Then, $\mathcal{C}_a^+=\{\text{X},\text{X},\text{Y},\text{Y}\}$. Defining $\mathcal{C}_a^+$ as a multiset allows to capture the number of times other areas ask for knowledge from area $a$.

Finally, let $\mathcal{C}_{aa'}^+\subset\mathcal{C}_a^+$ be a \textit{multisubset} consisting of courses in $\mathcal{C}_a^+$ whose area of study is $a'\in\mathcal{A}$. To measure the dependence of area $a'$ on area $a$, we define 
\begin{equation}\label{eqn:R_ij}
\begin{split}
R_{aa'}^{\mathcal{A}}&= \left(-\log\frac{|\mathcal{C}^+_{aa'}|}{|\mathcal{C}_a||\mathcal{C}_{a'}|}\right)^{-1}\\
&=\left(-\log\frac{\sum_{i,j\in\mathcal{G}}A_{ij}\delta(a_i,a)\delta(a_j,a')}{\sum_{i\in\mathcal{G}}\delta(a_i,a)\sum_{j\in\mathcal{G}}\delta(a_j,a')}\right)^{-1},
\end{split}
\end{equation}
where $\delta$ is the Kronecker delta: $\delta(x,y)=1$ if $x=y$ and $\delta(x,y)=0$ if $x\neq y$. The quantity under the logarithm has a probabilistic interpretation. The numerator is the total number of \textit{existing} links from courses in area $a$ to courses in area $a'$, and the denominator is the \textit{maximum possible} number of such links. So, if a course $X_a$ in area $a$ and a course $X_{a'}$ in area $a'$ are chosen at random, then $\exp(-1/R^{\mathcal{A}}_{aa'})$ is the the probability that there exists a link from $X_a$ to $X_{a'}$ in the CPN. 

From an information-theoretic point of view, the non-negative measure $R^{\mathcal{A}}_{aa'}\geq0$ is the inverse of the \textit{surprisal}~\cite{MacKay} (also called the \textit{Shannon information}) of the random event of observing a link from a randomly chosen $X_a\in\mathcal{C}_a$ to a randomly chosen $X_{a'}\in\mathcal{C}_{a'}$. The surprisal measures the level of ``surprise'' of occurring of that random event (existence of the link). As a result, the stronger the dependence of area $a'$ on area $a$ is, the smaller the surprisal (the larger the number of links from $a$ to $a'$), or, equivalently, the larger the measure $R^{\mathcal{A}}_{aa'}$. 

The dependence between academic divisions can be defined by analogy with (\ref{eqn:R_ij}). Namely, if $d,d'\in\mathcal{D}$ are two divisions, then the dependence of division $d'$ on division $d$ can be quantified by 
\begin{equation}
\begin{split}
R^{\mathcal{D}}_{dd'}&= \left(-\log\frac{|\mathcal{C}^+_{dd'}|}{|\mathcal{C}_d||\mathcal{C}_{d'}|}\right)^{-1}\\
&=\left(-\log\frac{\sum_{i,j\in\mathcal{G}}A_{ij}\delta(d_i,d)\delta(d_j,d')}{\sum_{i\in\mathcal{G}}\delta(d_i,d)\sum_{j\in\mathcal{G}}\delta(d_j,d')}\right)^{-1},
\end{split}
\end{equation}
which is the inverse of the information content of the event of observing a link from a random course offered by division $d$ to a random course offered by division $d'$. As with the areas of study, $R^{\mathcal{D}}_{dd'}\geq0$, and the larger $R^{\mathcal{D}}_{dd'}$ is, the stronger the dependence of $d'$ on $d$.  

\begin{figure}[h]
    \centering
    \includegraphics[width = \linewidth]{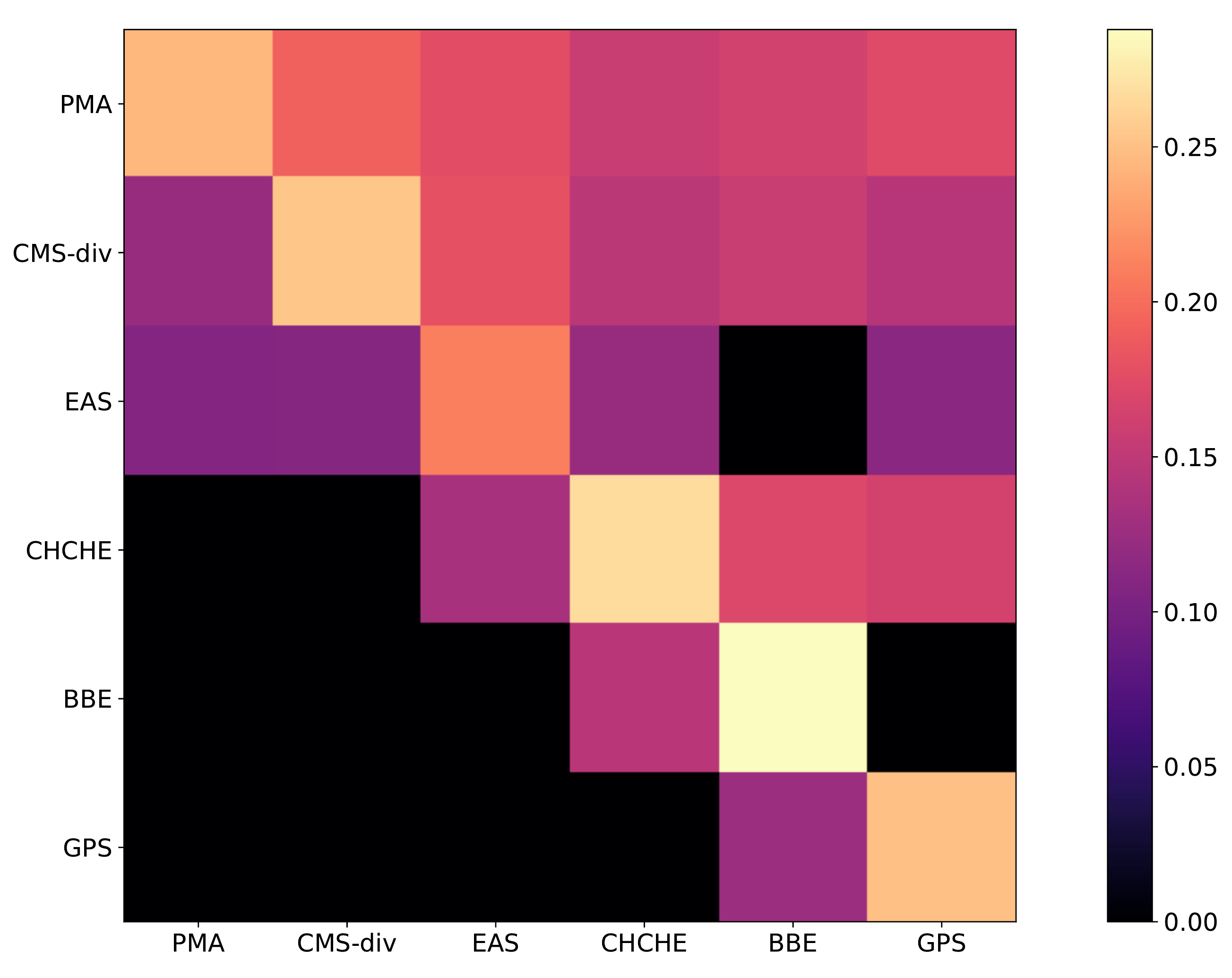}\\
    \vspace{3mm}
    \includegraphics[width = \linewidth]{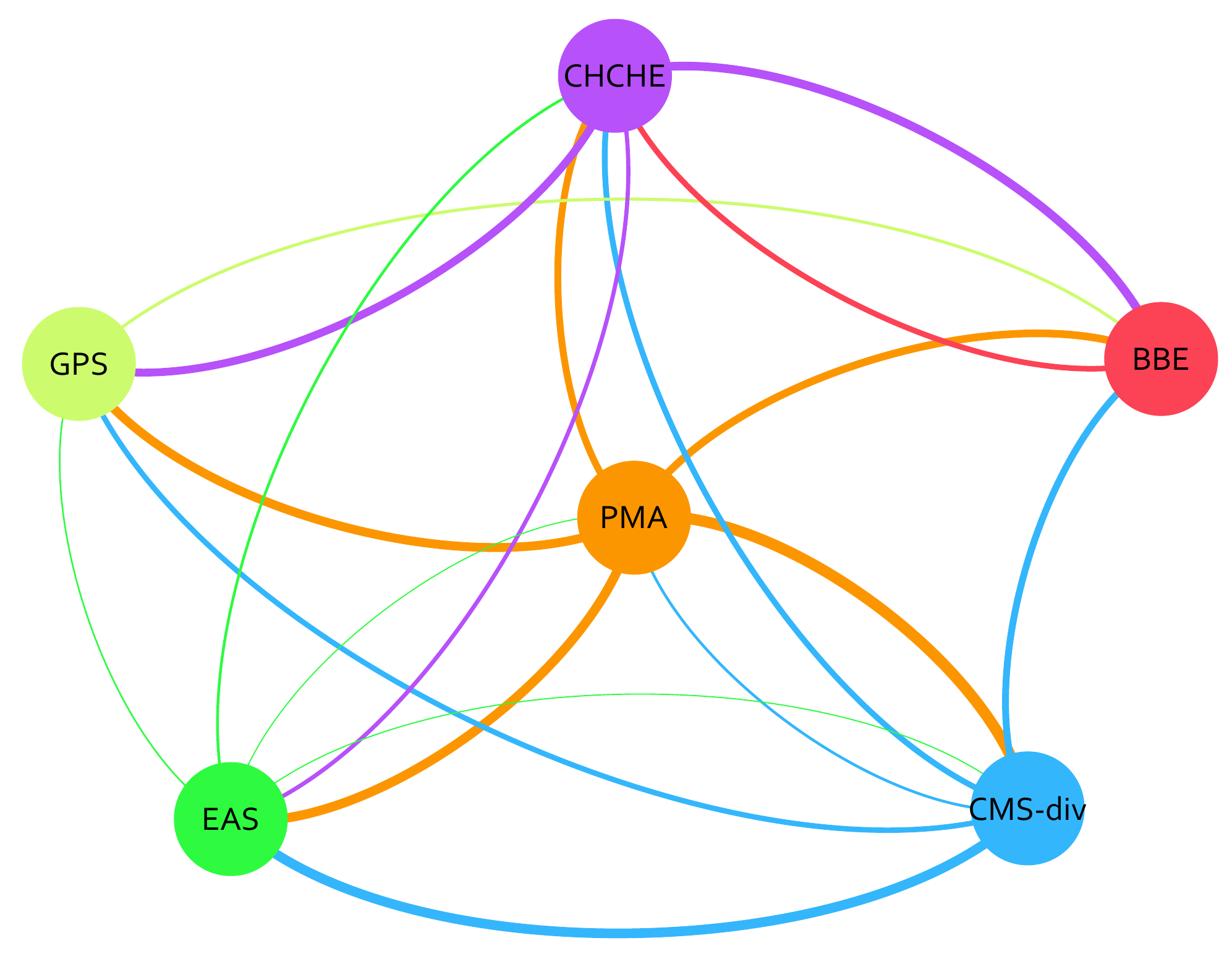}
    \caption{A heatmap visualization (top) and a network representation (bottom) of the division interdependence matrix $R^{\mathcal{D}}$ for the LCC $\mathcal{G}$ of the Caltech CPN. In the network representation, nodes represent divisions, each link is directed and starts from the node of the same color. The thickness of a link from $d$ to $d'$ is proportional to the value of $R^{\mathcal{D}}_{dd'}$. Self-links (from $d$ to $d$) are not shown.}
    \label{fig:Interdep_Div}
\end{figure}

Figure~\ref{fig:Interdep_Div} shows a heatmap visualization and a network representation of the division interdependence matrix $R^{\mathcal{D}}$ for the LCC of the Caltech CPN. 
As expected, the diagonal elements  $R^{\mathcal{D}}_{dd}$ have relatively large values for all $d\in\mathcal{D}$. This reflects the fact that courses offered by each division are  built on top of each other, and more elementary division courses serve as prerequisites for more advanced division courses. The largest value of $R^\mathcal{D}_{dd}$ is for BBE, whose courses are all on biological subjects and are strongly \textit{intradependent}. 

The two most \textit{influential} divisions, in the sense that all other divisions depend on them, are PMA and CMS-div. These divisions offer courses on mathematics, physics, and computer science, which serve as crucial prerequisites for science and engineering courses offered by other divisions. Interestingly, the dependence of CMS-div on PMA is stronger than the dependence of PMA on CMS-div, which indicates that applied mathematics draws from pure mathematics more than the other way around. The largest \textit{clique} in the division interdependence network in Fig.~\ref{fig:Interdep_Div}, \ie the largest subset of fully interconnected nodes, is PMA, CMS-div, EAS. There is a great flow of knowledge between these three divisions, especially from PMA to CMS-div and EAS, and from CMS-div to EAS. 

The most \textit{interdisciplinary} divisions, in the sense that they require knowledge from and depend on most other divisions, are CHCHE, BBE, and GPS. These divisions have a large number of incoming links, contrary to the most influential divisions that have a large number of outgoing links. Being centered around biological, chemical, and geological sciences, CHCHE, BBE, and GPS depend on mathematical methods, computational tools, and physical theories taught in courses offered by other divisions. The converse, however, is not true: PMA and CMS-div do not have incoming links from CHCHE, GPS or BBE. 

\begin{figure}[h]
    \centering
    \includegraphics[width = \linewidth]{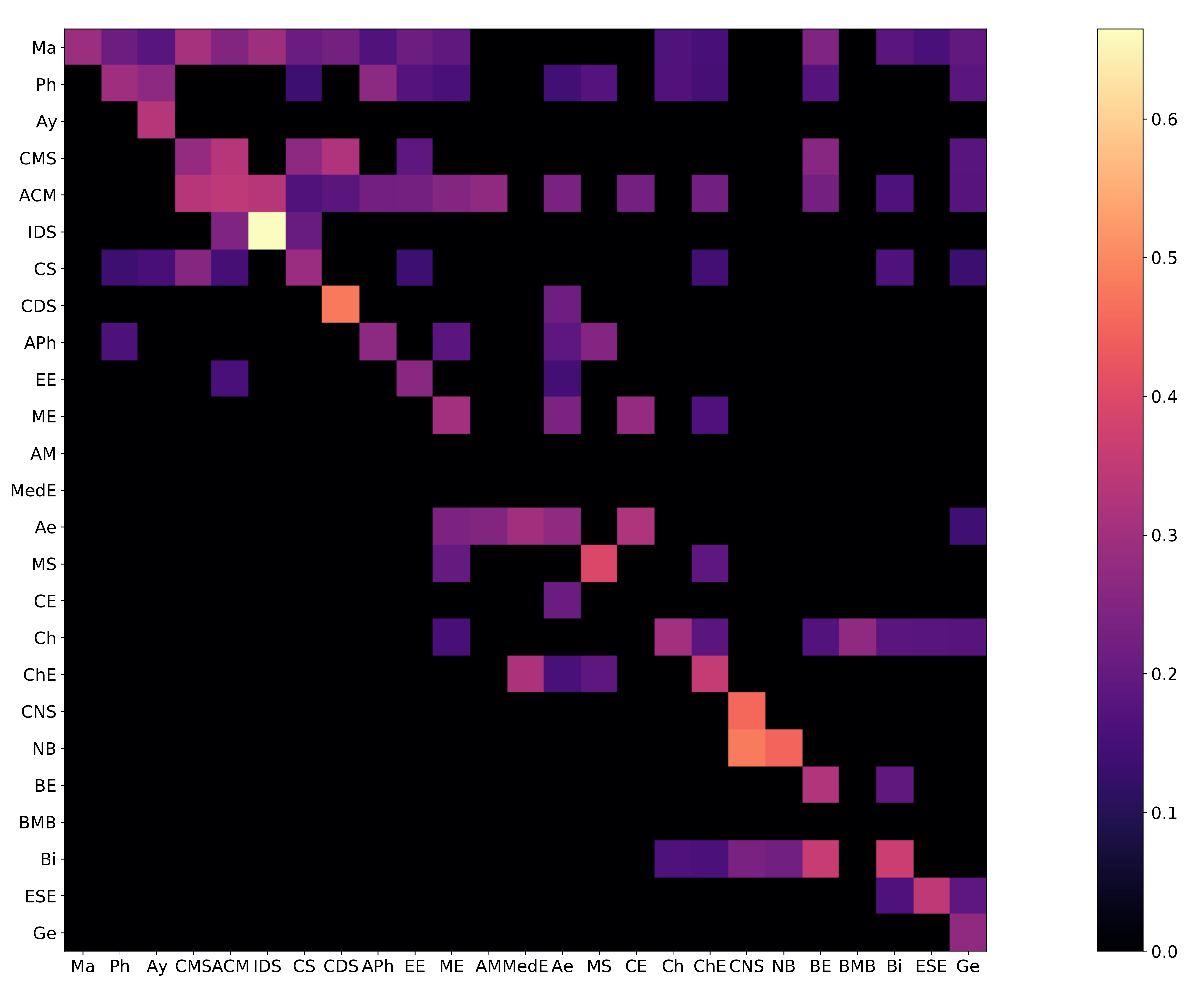}\\
    \vspace{3mm}
    \includegraphics[width = \linewidth]{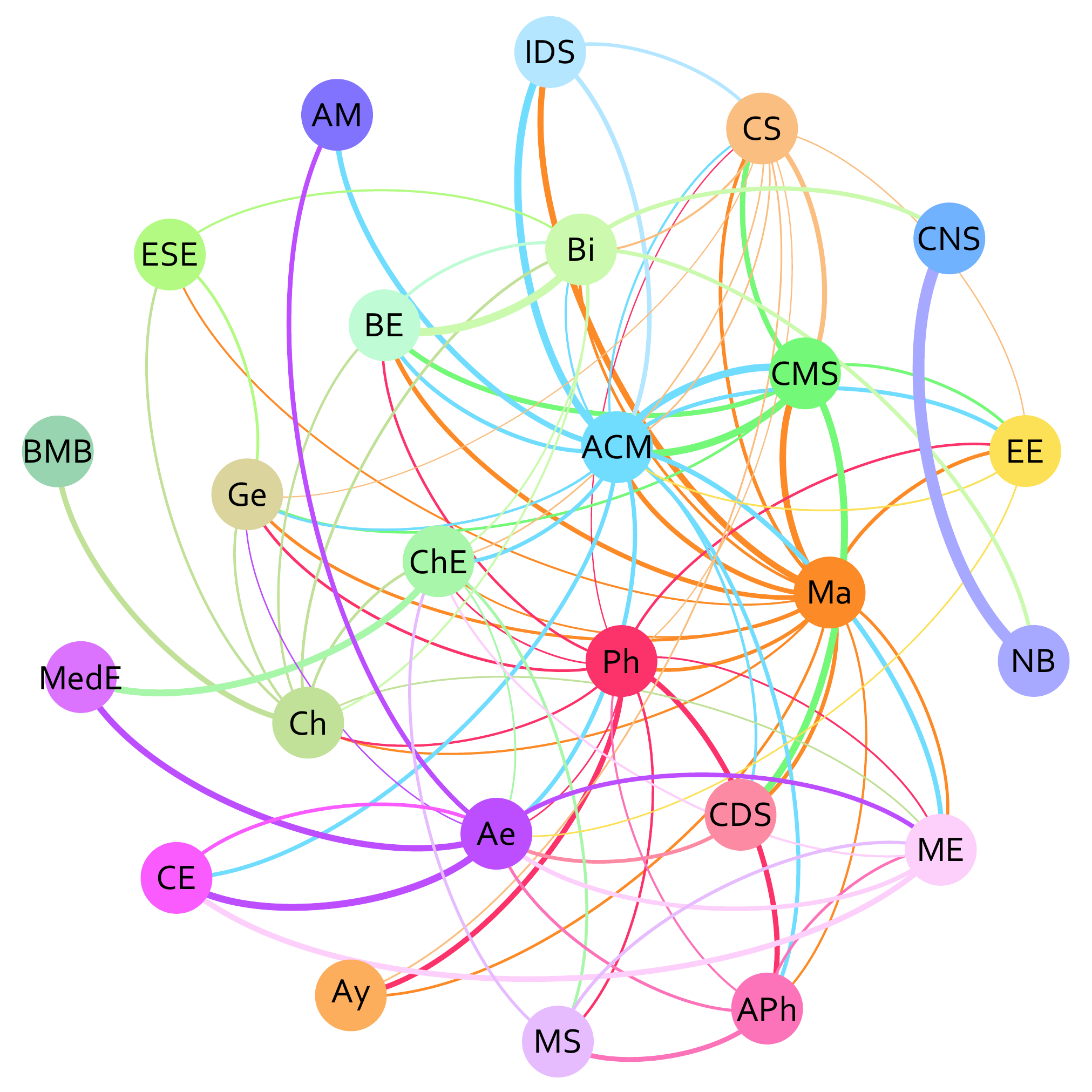}
    \caption{A heatmap visualization (top) and a network representation (bottom) of the area-of-study interdependence matrix $R^{\mathcal{A}}$ for the LCC $\mathcal{G}$ of the Caltech CPN. In the network representation, nodes represent areas of study, each link is directed and starts from the node of the same color. The thickness if a link from $a$ to $a'$ is proportional to the value of $R_{aa'}^{\mathcal{A}}$. Self-links (from $a$ to $a$) are not shown. For visual clarity, a large scale visualization of this network is shown in Fig.~\ref{fig:interdep_big} in the Appendix.}
    \label{fig:Interdep_small}
\end{figure}

Extending the interdependence analysis from the division level to the level of areas of study, Fig.~\ref{fig:Interdep_small} shows a heatmap visualization and a network representation of matrix $R^{\mathcal{A}}$ for the LCC of the Caltech CPN. It can be viewed as a fine-grained version of Fig.~\ref{fig:Interdep_Div}.
As in the case of divisions, the diagonal elements $R_{aa}^{\mathcal{A}}$ have relatively large values for most $a\in\mathcal{A}$ for a similar reason. Namely, all courses offered in a certain area of study naturally form sequences, where more advanced courses have more elementary courses as prerequisites; this makes the \textit{intra}-dependence of the area of study strong. The largest value of $R_{aa}^{\mathcal{A}}$ is for IDS (Information and Data Sciences), the newest option introduced at Caltech in 2018, which has a small number of interconnected courses. On the other extreme, a few areas of study, such as Medical Engineering (MedE) and Biochemistry and Molecular Biophysics (BMB), have zero $R_{aa}^{\mathcal{A}}$. These areas are originated at the intersection of more traditional areas, and have a small number of courses, whose prerequisites come from the said traditional areas. 

The most \textit{influential} areas of study in the Caltech CPN with the biggest impact on other areas are Mathematics (Ma), Applied and Computational Mathematics (ACM), Physics (Ph), and Computer Science (CS). All these areas belong to the two most influential divisions, PMA and CMS-div. Outgoing links from the nodes representing these areas reach out to almost all other nodes and form the backbone of the area network in Fig.~\ref{fig:Interdep_small}. The Ma node plays a unique role in the network: it has the largest number of outgoing links, and yet it is the only node that does not have any incoming links. This reflects the fact that many areas depend on mathematics, but mathematics itself does not depend on other areas, at least in the curriculum. The ACM node, an ``applied cousin'' of Ma, is also very important for other areas, especially for engineering and physical sciences, but, unlike its ``pure cousin,'' does have a few incoming links. Of particular interest is the interaction between Ph and CS. The main reason behind this interaction is the recent rise of quantum computing, which led to new physics and computer science courses. We expect that, as quantum computing becomes more developed and accessible, the interaction between Ph and CS will become even stronger. The largest clique in the area interdependence network is ACM, CMS and CS, which all belong to CMS-div.

Among the most \textit{interdisciplinary} areas of study, \ie areas with a large number of incoming links, are Aerospace (Ae), Chemical Engineering (ChE), and Geology (Ge).  Interestingly, while Ae and ChE also have outgoing links and, thus, have influence on other areas of study, Ge, one of the largest areas at Caltech, has no influence on any other areas. 

It is worth mentioning several strong interactions between specialized but related areas. For example, Aerospace (Ae) has a strong impact on Civil Engineering (CE), Mechanical Engineering (ME), Applied Mechanics (AM), and Medical Engineering (MedE). The engineering areas have a significant dependence on Ae because the latter offers important courses on theoretical and computational fluid mechanics which serve as prerequisites for CE, ME, AM, and MedE courses. Other notable interactions include a very strong dependence of Computation \& Neural Systems (CNS) on Neurobiology (NB) and the impact of Biology (Bi) on Biological Engineering (BE), CNS and NB. 

Table~\ref{tab:PB_areas} lists the top areas of study with respect to the out-degree, in-degree, PageRank and betweenness centralities computed for the area network in Fig.~\ref{fig:Interdep_small} with adjacency matrix $A^{\mathcal{A}}$ defined as follows:
\begin{equation}
A^{\mathcal{A}}_{aa'}=\begin{cases}
0, & \mbox{if } a=a' \mbox{ (no self-loops)},\\
0, & \mbox{if } R_{aa'}^{\mathcal{A}}=0,\\
1, & \mbox{if } a\neq a' \mbox{ and } R_{aa'}^{\mathcal{A}}>0.
\end{cases}
\end{equation} 
Interdisciplinary engineering areas dominate the in-degree ranking. The out-degree and PageRank rankings are very similar (like in Section~\ref{sec:CentralityMeasures}) and are dominated by mathematical and computational areas, the most influential areas in the network. The top five betweenness areas come from four different divisions and they serve as  critical bridges in transferring knowledge between various areas of study. The fact that ACM is the only area that appears in the top PageRank and betweenness rankings highlights its prominent role in the entire curriculum. 
\begin{table}[h]
	\centering
	\begin{tabular}{c|c|c|c|c}
		\multicolumn{5}{c}{\textsc{\textcolor{black}{Top 5 Areas of Study}}} \\
		\hline
		& \textsc{\textcolor{black}{In-Degree}} & \textsc{\textcolor{black}{Out-Degree}}&\textsc{\textcolor{black}{PageRank}}  & \textsc{\textcolor{black}{Betweenness}}\\
		\hline
		\textcolor{black}{1} & Ae & Ma  & Ma & Bi\\
		\hline
		\textcolor{black}{2} & ChE & ACM  & ACM & ACM\\
		\hline
		\textcolor{black}{3} & Ge & Ph  & CS & Ae\\
		\hline
		\textcolor{black}{4} & ME & CS  & Ph & ChE\\
		\hline
		\textcolor{black}{5} & BE & Ch & CMS & Ch
	\end{tabular}
	\caption{The top 5 areas of study with respect to the in-degree, out-degree, PageRank and betweenness.}
	\label{tab:PB_areas}
\end{table}

The interdependence analysis shows that there is a strong flow of knowledge from mathematical, physical, and computational sciences to applied and engineering sciences, or more generally, from more theoretical to more applied fields. On the other hand, the flow of knowledge in the reverse direction is rather weak at both division and area-of-study levels. This observation manifests itself in the fact that the interdependence matrices $R^{\mathcal{D}}$ and $R^{\mathcal{A}}$ visualized in Fig.~\ref{fig:Interdep_Div} and Fig.~\ref{fig:Interdep_small} are approximately upper triangular. Although there are important ``feedback'' links,  such as APh $\rightarrow$ Ph, EE $\rightarrow$ ACM, where more theoretical areas draw knowledge form more applied areas, the number of such links is rather small. This suggests that to increase the \textit{inter}dependence of the Caltech curriculum, to make it more balanced and interdisciplinary, new theoretical courses with applied prerequisites are needed. These applied prerequisites could provide important application domains for concepts and methods taught in theoretical courses, thereby giving rise to a healthy interaction between theory and applications in the university curriculum.

\section{Summary and Discussion}
\label{sec:Conclusions}

In this paper, we introduced course-prerequisite networks (CPNs), proposed a general network-science-based framework for their analysis, and illustrated it with a CPN constructed from courses taught at the California Institute of Technology. Conceptually, a CPN represents the flow of knowledge in a university curriculum. This network can be used for curriculum visualization, finding important insights about courses, helping students to navigate the curriculum, and helping faculty and administrators to optimize it.   

First, we focused on the identification of important courses in a given CPN. Since ``importance'' can be defined differently, we considered four different centrality metrics that measure different aspects of importance. The in-degree measures how specialized a course is: the larger the in-degree, the more prerequisites the course has, and the more specialized it is. The top in-degree courses require a lot of preparation, and students should be advised to take these courses towards the end of their studies. The out-degree measures how fundamental a course is: the larger the out-degree, the more postrequisites the course has, and the more fundamental it is. The top out-degree courses are usually the most fundamental courses in mathematics, natural sciences, and computer science, and students should aim at taking these courses as soon as possible. The in- and out-degrees are negatively correlated, and the absolute value of the Pearson correlation coefficient between in- and out-degrees of nodes  can be used a measure of the structural ``gap'' between fundamental and specialized courses in the CPN. 

The out-degree is a more suitable measure of course importance than the in-degree, since the out-degree of a course is determined by other courses, while its in-degree is determined by the course itself. Similar to the out-degree, the PageRank centrality measures how fundamental a course is, but it favors more introductory courses. Unlike the out-degree, PageRank takes into account not only the number of postrequisites, but also their importance. Finally, betweenness identifies intermediate-level courses that serve is critical bridges between less and more advanced courses. For the Caltech CPN, we found that PageRank and betweenness are positively correlated, and courses which are important with respect to one of the measures tend to be also important with respect to the other. Courses that are ranked highly by both PageRank and betweenness play a vital role in the university curriculum. Students should be advised to pay special attention to these courses. Maximizing the quality of these courses is the key step in improving the whole curriculum. Administrators should, therefore, prioritize enhancing the quality of courses with high PageRank and betweenness scores by ensuring that these courses are taught well (for example, by hiring dedicated teaching faculty) and students are provided with all necessary resources, such as office hours, recitation sessions, and tutoring support. 

Next, we introduced the topological stratification of a CPN, which is the ordered partition of the CPN nodes into disjoint subsets, or strata, of topologically equivalent nodes. The topological stratification of a CPN allows to view the hierarchy of the whole university curriculum with respect to the level of advancement, starting from the most elementary undergraduate courses and building up to the most advanced courses at the graduate level. Students can use these strata to make more informed choices about what courses to take in order to obtain a deeper understanding of their area of study. Furthermore, topological stratification unveils hidden prerequisites, \ie courses that are not formal prerequisites for a given course, but would be beneficial if completed before taking the course. For each course in the curriculum, a group of faculty and students can examine the list of its identified hidden prerequisites and make a decision on whether to make them formal prerequisites (or list them as highly recommended).  

A hierarchical structure (\ref{eq:hierarchicalstructure}) on a CPN induced by its topological stratification can be reduced on any specific major or area of study. A group of faculty and students can examine the resulting reduced hierarchy (\ref{eq:hierarchicalstructureM}) and decide whether it indeed provides an comprehensive learning path for obtaining a thorough understanding of the major. The reduced hierarchy can be also used by undergraduate students as a navigation tool and by their academic advisors as a trusted reference for helping students to choose courses.

Finally, the interdependence analysis provides a bird's eye view of the whole CPN at the division level (crane's eye view) and the area-of-study level (owl's eye view). It allows to quantify the strength of flow of knowledge from one university division or area of study to another and identify the most intradependent, influential, and interdisciplinary divisions and areas of study. The university administration can use the results of interdependence analysis for optimal allocation ot teaching resources to ensure that divisions and departments, which offer courses that many other divisions and departments depend on, do have adequate means for providing their crucial teaching service. The interdependence networks can help faculty to assess their interactions with other divisions and areas of study and increase these interactions by creating new interdisciplinary courses. Finally, students can also use the interdependence networks to discover how they can branch out into new areas of study. To this end, students should find areas that draw knowledge from their current area of study and take course in those areas. This may lead to the development of new academic interests and interdisciplinary studies. 

It is important to highlight that although the proposed methodology for CPN data analysis was illustrated with the Caltech CPN, it is by no means specific to Caltech. In fact, it can be used -- in a straightforward way and in a complete analogy with the considered example -- for the curriculum analysis of essentially any university. To this end, the following CPN data is required: 
\begin{itemize}
	\item[(a)] List if nodes, representing courses. Depending on the goals of the analysis, this list could be limited to only undergraduate courses, only graduate courses, or, like in the considered example, it could include all courses.
	\item[(b)] List of directed links, representing prerequisite relationships. This list can be extracted from the university catalog that contains the description of courses.
	\item[(c)] Partition of nodes into academic divisions (or other administrative units, such as departments, schools, and colleges).
	\item[(d)] Partition of nodes into areas of study, defined by the degree options.
\end{itemize}

All real networks evolve with time, and so do course-prerequisite networks. Although the speed of evolution of CPNs is relatively low compared to many other networks, such as the world wide web, citation networks, and online social networks, adding new courses, removing old ones, and establishing new prerequisite relationships between courses may eventually change the structure and quantitative properties of a CPN. To keep up with the CPN evolution and have updated results, the described CPN analysis can be redone on a regular basis, for instance every three to five years, depending on how quickly the CPN changes.

To summarize, a CPN is an indispensable tool for summarizing, visualizing, and analyzing an academic curriculum. It can help to better understand and revamp the curriculum, detect important courses, improve existing and create new courses, meaningfully allocate teaching resources, increase interdisciplinary interactions between various university units, and enhance the overall student learning experience.    

\begin{acknowledgments}
We thank Gloria Brewster and Kimberley Mawhinney for providing us with a database of Caltech courses, Eric V. Smith for the valuable help with the toposort module, Justin Bois for useful feedback on the first draft of the paper, and Liza Bradulina and Sophia Zueva for stimulating discussions. We also thank the anonymous reviewers for useful comments and suggestions. This work was supported by the Carver Mead Discovery Grant and the Information Science and Technology (IST) initiative at Caltech.
\end{acknowledgments}

\onecolumngrid
\newpage

\section*{Appendix: Caltech CPN Visualizations}

\begin{figure}[h]
	\centering
	\includegraphics[width = \textwidth]{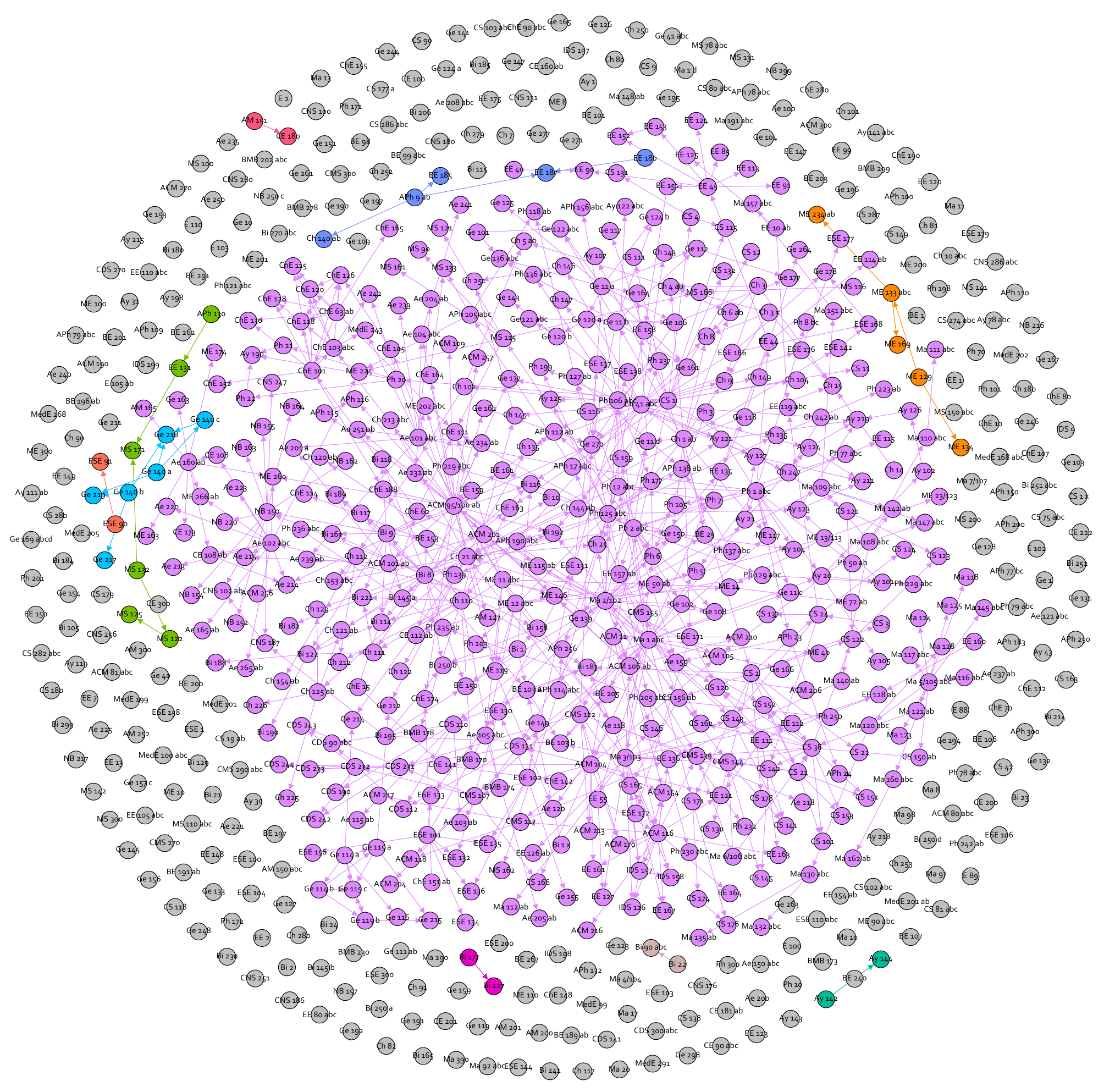}
	\caption{The 2021-2022 Caltech CPN with course names as node labels. Gray nodes are isolated. Nodes in color other than gray represent connected components. The network has $771$ nodes and $772$ links.}
	\label{fig:Caltech_CPN_full_labels}
\end{figure}

\newpage

\begin{figure}[h]
	\centering
	\includegraphics[width = \textwidth]{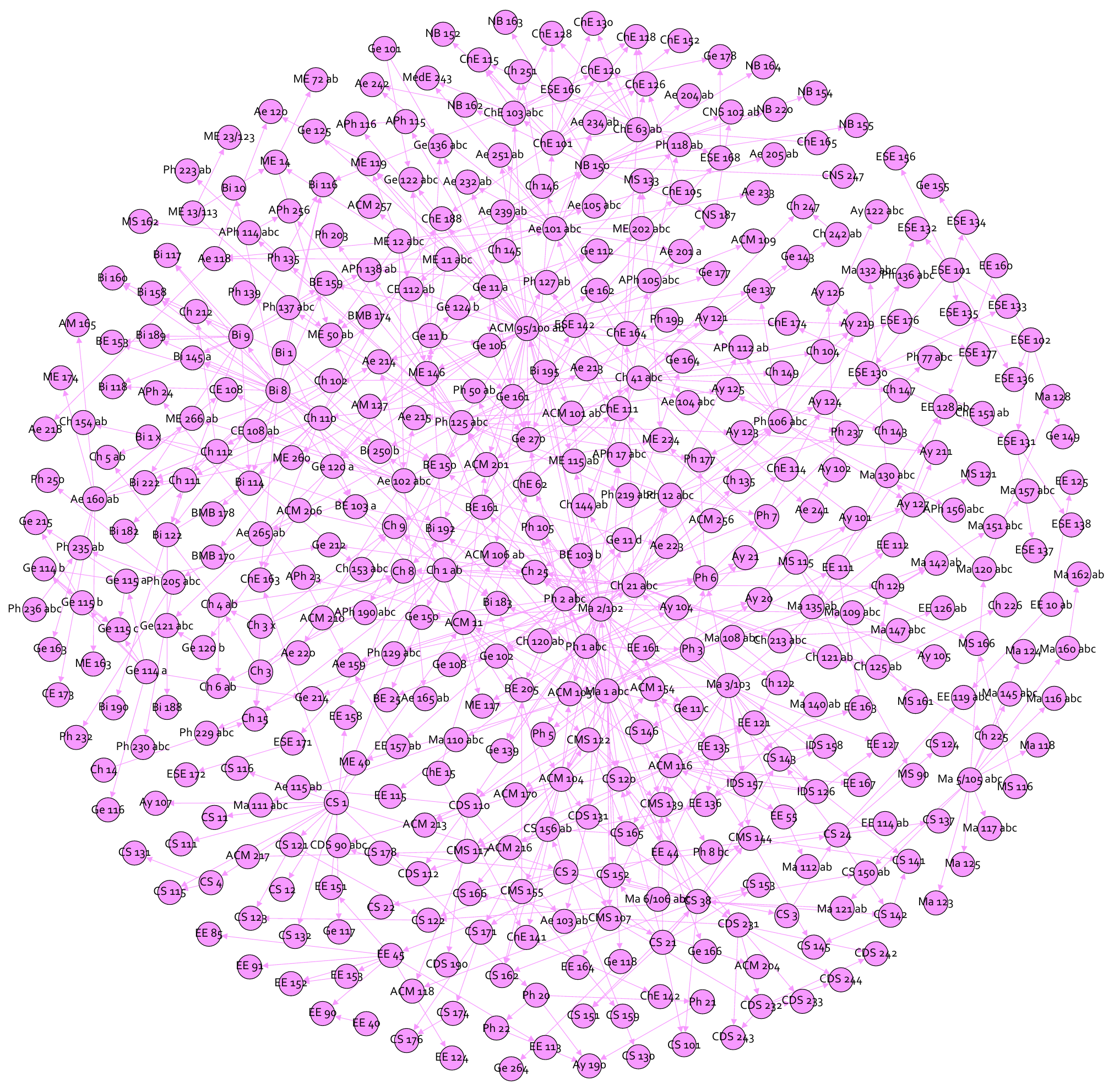}
	\caption{The largest connected component (LCC) $\mathcal{G}$ of the Caltech CPN shown in full in Fig.~\ref{fig:Caltech_CPN_full_labels}. The LCC has $n=436$ nodes ($57$\% of all nodes) and  $m=747$ links ($97$\% of all links).}
	\label{fig:Caltech_LCC}
\end{figure}

\begin{figure}[h]
	\centering
	\includegraphics[width = \linewidth]{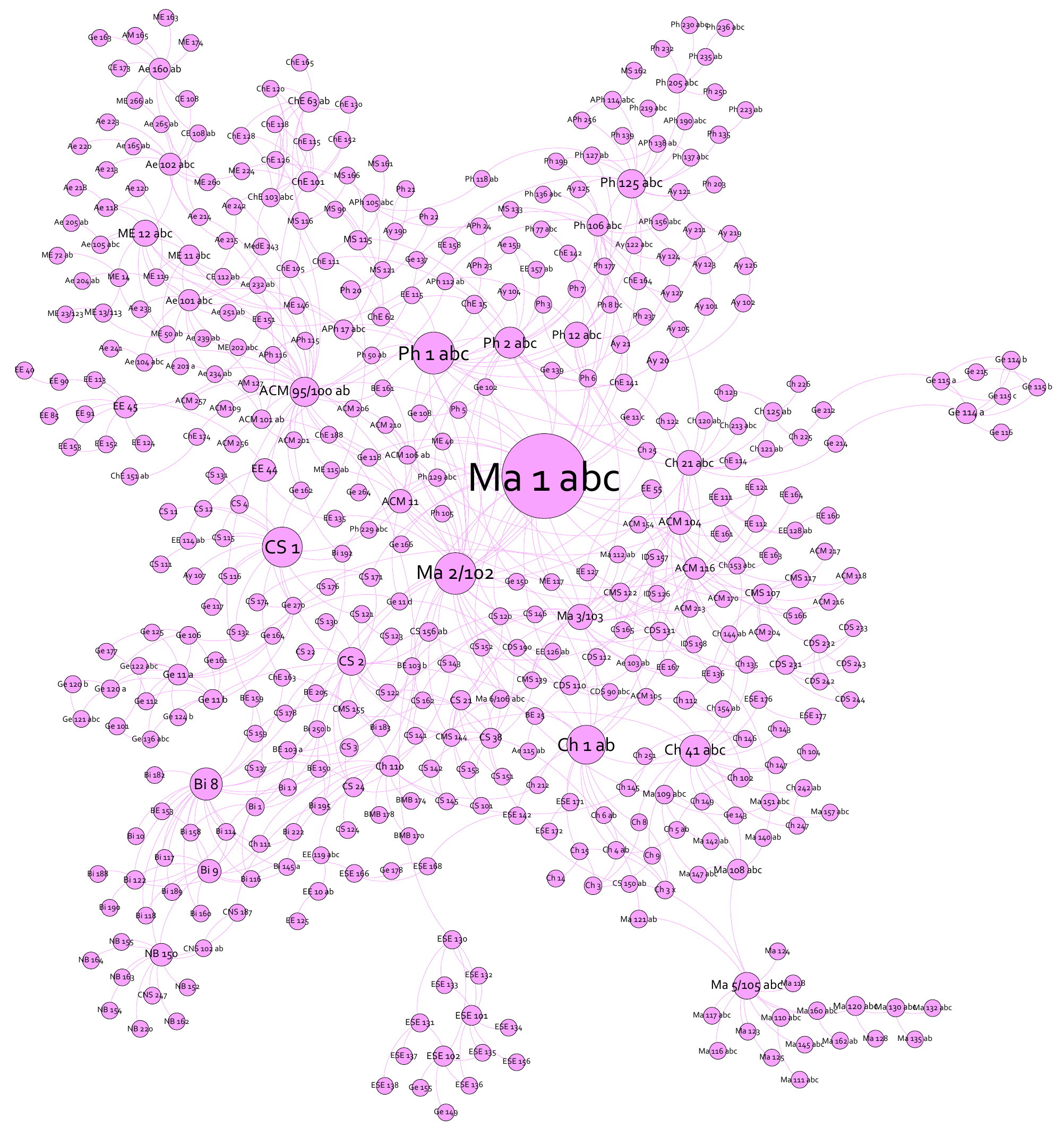}
	\caption{The LCC $\mathcal{G}$ of the Caltech CPN visualized with node size proportional to node's PageRank centrality: the larger the node, the higher its PageRank.}
	\label{fig:LCC_pr_full}
\end{figure}

\begin{figure}[h]
	\centering
	\includegraphics[width = 0.95\linewidth]{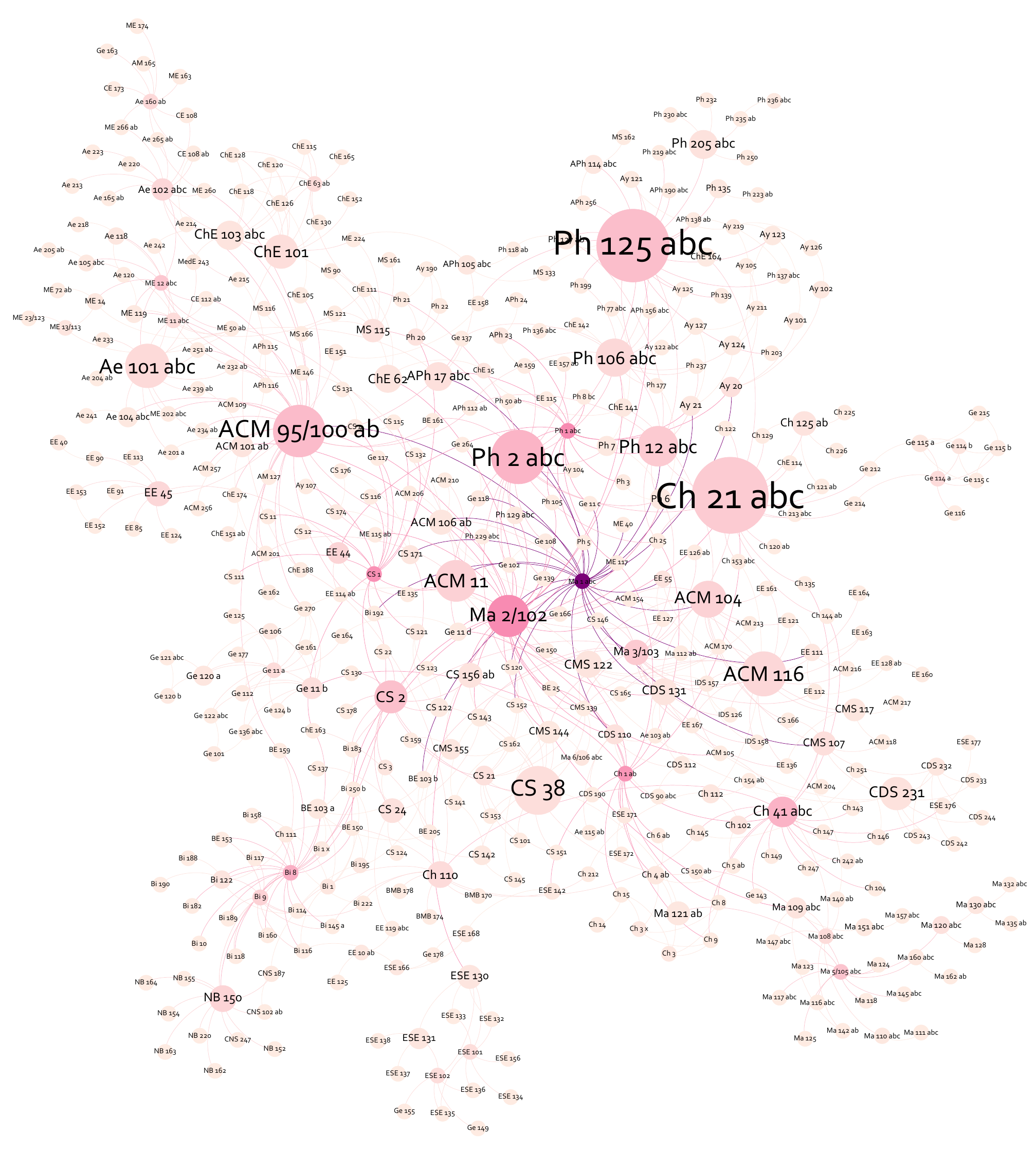}
	\caption{The LCC $\mathcal{G}$ of the Caltech CPN visualized with node size proportional to node's betweenness and color intensity proportional to node's PageRank. The larger and darker the node, the higher its betweenness and PageRank, the more important it is. Links are directed and start at the node of the same color.}
	\label{fig:LCC_pr_bet_full}
\end{figure}

\begin{table}[H]
    \centering
    \begin{tabular}{|c|c|}
    \multicolumn{2}{c}{\textsc{Topological Stratification of the LCC} $\mathcal{G}$ \textsc{of the Caltech CPN}}\\
    \hline
        $\mathcal{S}_1$
         & CS 1, Ma 1\\
         \hline
    $\mathcal{S}_2$ & ACM 11, Ch 1, ChE 15, Ma 2\\
    \hline
    $\mathcal{S}_3$ & ACM 104, Bi 8, CS 2, ChE 62, ChE 63, Ge 11 ab, Ma 5, Ph 1\\
    \hline
    \multirow{2}{*}{$\mathcal{S}_4$} & ACM 95/100, APh 17, Ay 20, Bi 1, Bi 1 x, Bi 9, CDS 131, CMS 107, CMS 122, CS 21, CS 3, Ch 41, ChE 101,\\
    & EE 44, EE 55, ESE 101, ESE 102, Ge 106, Ge 114 a, ME 11, ME 12, Ma 108, Ma 121, Ma 3, Ma 6, Ph 12, Ph 2, Ph 3\\
    \hline
    \multirow{3}{*}{$\mathcal{S}_5$} & ACM 116, Ae 101, Ae 102, Ae 160, Ay 101, Ay 102, Ay 21, Be 103 a, Bi 122, Bi 195, CDS 110, CDS 231, CS 121, \\
    & CS 156, CS 171, CS 24, CS 38, Ch 102, Ch 110, Ch 21, Ch 3, ChE 103, ChE 105, EE 111, EE 45, ESE 130, ESE 142,\\
    & ESE 166, Ge 114 b, Ge 120 a, ME 13, Ma 109, Ma 120, NB 150, Ph 106, Ph 125, Ph 20, Ph 6\\
    \hline
    \multirow{7}{*}{$\mathcal{S}_6$} & ACM 101, ACM 106, APh 105, APh 114, APh 115, APh 23, Ae 104, Ae 105, Ae 118, Ay 121, Ay 123, Ay 124, Ay 126,\\
    & Ay 127, BE 103 b, CDS 112, CDS 232, CE 108, CMS 117, CMS 144, CMS 155, CNS 187, CS 122, CS 142, CS 143,\\
    & CS 174, CS 4, Ch 111, Ch 112, Ch 125, Ch 14, Ch 145, Ch 146, Ch 242, Ch 4, ChE 126, ChE 141, ChE 151, ChE 164,\\
    & EE 10, EE 112, EE 151, EE 153, EE 157, EE 160, EE 40, ESE 131, ESE 168, ESE 171, ESE 176, Ge 101, Ge 11 d,\\
    & Ge 112, Ge 115 ab, Ge 118, Ge 120 b, Ge 161, Ge 162, IDS 157, ME 119, ME 14, MS 115, MS 116, Ma 110, Ma 130,\\
    & Ma 140, Ma 151, Ma 160, Ph 127, Ph 129, Ph 135, Ph 137, Ph 205, Ph 21, Ph 22, Ph 236, Ph 7\\
    \hline
    \multirow{25}{*}{$\mathcal{S}_7$} & ACM 105, ACM 109, ACM 118, ACM 154, ACM 170, ACM 201, ACM 204, ACM 206, ACM 210, ACM 213, ACM 216,\\
    & ACM 217, ACM 256, ACM 257, AM 127, AM 165, APh 112, APh 116, APh 138, APh 156, APh 190, APh 24, APh 256,\\
    & Ae 103, Ae 115, Ae 120, Ae 159, Ae 165, Ae 201, Ae 204, Ae 205, Ae 213, Ae 214, Ae 215, Ae 218, Ae 220, Ae 223,\\
    & Ae 232, Ae 233, Ae 234, Ae 239, Ae 241, Ae 242, Ae 251, Ae 265, Ay 104, Ay 105, Ay 107, Ay 122, Ay 125, Ay 190,\\
    & BE 150, BE 153, BE 159, BE 161, BE 205, BE 25, BMB 170, BMB 174, BMB 178, Bi 10, Bi 114, Bi 116, Bi 117,\\
    & Bi 118, Bi 145, Bi 158, Bi 160, Bi 182, Bi 183, Bi 188, Bi 189, Bi 190, Bi 192, Bi 222, Bi 250, CDS 190, CDS 233,\\
    & CDS 242, CDS 243, CDS 244, CDS 90, CE 108, CE 112, CE 173, CMS 139, CNS 102, CNS 247, CS 101, CS 11, CS 111,\\
    & CS 115, CS 116, CS 12, CS 120, CS 123, CS 124, CS 130, CS 131, CS 132, CS 137, CS 141, CS 145, CS 146, CS 150,\\
    & CS 151, CS 152, CS 153, CS 159, CS 162, CS 165, CS 166, CS 176, CS 178, CS 22, Ch 104, Ch 120, Ch 121, Ch 122,\\
    & Ch 129, Ch 135, Ch 143, Ch 144, Ch 147, Ch 149, Ch 15, Ch 153, Ch 154, Ch 212, Ch 213, Ch 225, Ch 226, Ch 247,\\
    & Ch 25, Ch 251, Ch 5, Ch 6, Ch 8, Ch, 9, ChE 111, ChE 114, ChE 115, ChE 118, ChE 120, ChE 128, ChE 130, ChE 142,\\
    & ChE 152, ChE 163, ChE 165, ChE 174, ChE 188, EE 113, EE 114, EE 115, EE 119, EE 121, EE 124, EE 125, EE 126,\\
    & EE 127, EE 128, EE 135, EE 136, EE 152, EE 158, EE 161, EE 163, EE 164, EE 167, EE 85, EE 90, EE 91, ESE 132,\\
    & ESE 133, ESE 134, ESE 135, ESE 136, ESE 137, ESE 138, ESE 156, ESE 172, ESE 177, Ge 102, Ge 11 c, Ge 115 c,\\
    & Ge 116, Ge 117, Ge 121, Ge 122, Ge 124, Ge 125, Ge 136, Ge 137, Ge 139, Ge 143, Ge 149, Ge 150, Ge 155, Ge 163,\\
    & Ge 164, Ge 166, Ge 177, Ge 178, Ge 212, Ge 214, Ge 215, Ge 264, Ge 270, IDS 126, IDS 158, ME 115, ME 117,\\
    & ME 146, ME 163, ME 174, ME 202, ME 224, ME 23, ME 260, ME 266, ME 40, ME 50, ME 72, MS 121, MS 133,\\
    & MS 161, MS 162, MS 166, MS 90, Ma 111, Ma 112, Ma 116, Ma 117, Ma 118, Ma 123, Ma 124, Ma 125, Ma 128,\\
    & Ma 132, Ma 135, Ma 142, Ma 145, Ma 147, Ma 157, Ma 162, MedE 243, NB 152, NB 154, NB 155, NB 162, NB 163,\\
    & NB 164, NB 220, Ph 105, Ph 118, Ph 136, Ph 139, Ph 177, Ph 199, Ph 203, Ph 219, Ph 223, Ph 229, Ph 230, Ph 232,\\
    & Ph 235, Ph 237, Ph 250, Ph 5, Ph 50, Ph 77, Ph 8\\
    \hline
    \end{tabular}
    \caption{Topological stratification (\ref{eq:topstrat}) of the LCC $\mathcal{G}$ of the Caltech CPN. Each stratum contains topologically equivalent courses, starting with the most fundamental $(\mathcal{S}_1)$ and leading to the most advanced $(\mathcal{S}_7)$.}
    \label{tab:Stratification}
\end{table}

\begin{figure}[h]
    \centering
    \includegraphics[width = \linewidth]{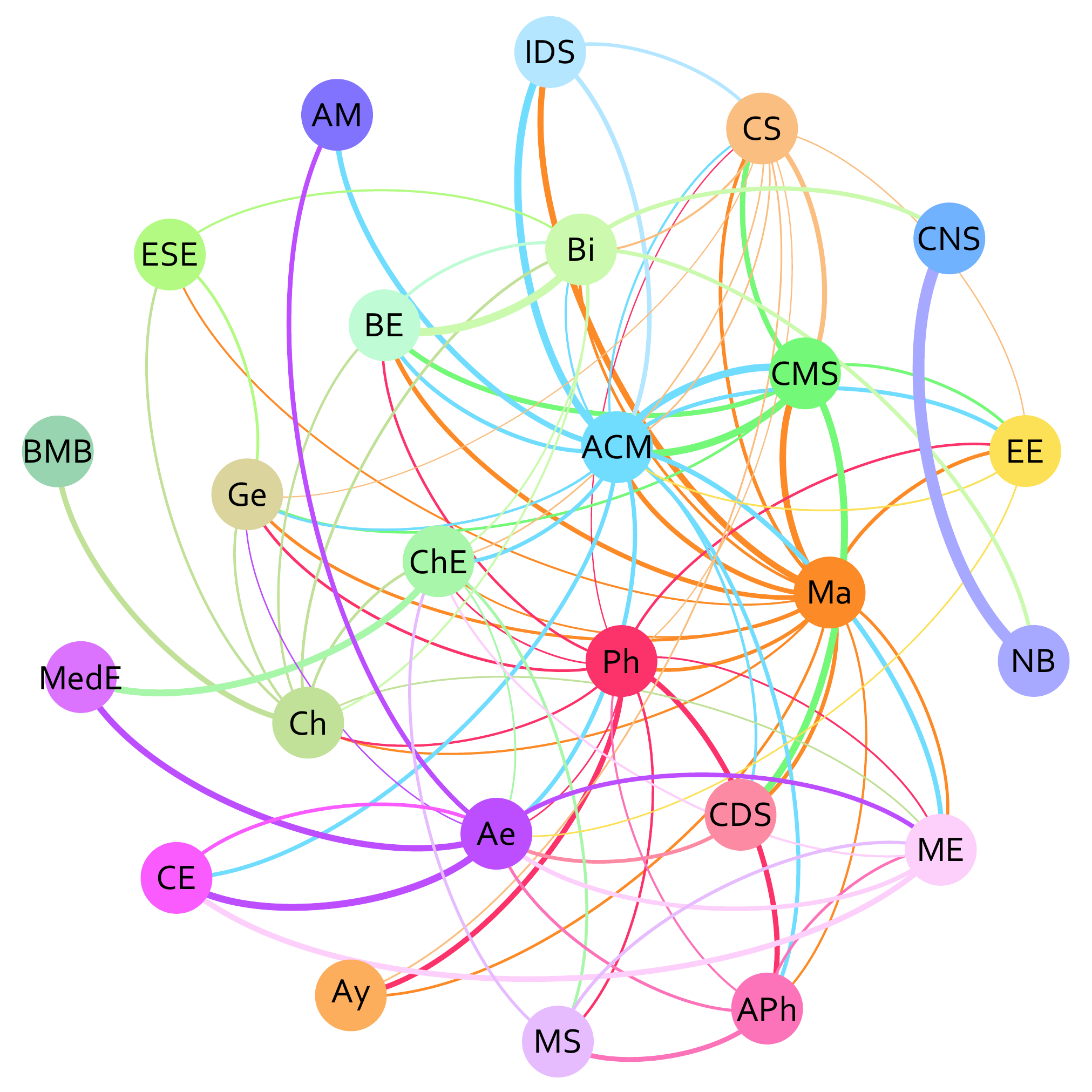}
    \caption{A network representation of the area-of-study interdependence matrix $R^{\mathcal{A}}$ for the LCC $\mathcal{G}$ of the Caltech CPN. Nodes represent areas of study, each link is directed and starts from the node of the same color. The thickness if a link from $a$ to $a'$ is proportional to the value of $R_{aa'}^{\mathcal{A}}$. Self-links (from $a$ to $a$) are not shown.}
    \label{fig:interdep_big}
\end{figure}

\twocolumngrid


\begin{thebibliography}{99}
	
	\bibitem{Akbas2015}
	Akbas, M.I., Basavaraj, P., Georgiopoulos, M., Garibay, {\"O}., \& Garibay., I. Curriculum GPS: an adaptive curriculum generation and planning system. \textit{Interservice/Industry Training, Simulation, and Education Conference}, (2015).
	
    \bibitem{Slim2014}
    Slim A., Heileman G.L., Kozlick J. \& Abdallah C.T. Employing Markov networks on curriculum graphs to predict student performance.  \textit{Proceedings of the 13th International Conference on Machine Learning and Applications}, IEEE, 415-418, (2014). 
    
    \bibitem{Meghanathan2017}
    Meghanathan N. Curriculum network graph: relative contribution of courses. \textit{International Journal of Network Science} \textbf{1}(3), 223-247 (2017).
    
    \bibitem{Blas2021}
    Simon Blas, S.C., Gonzalez, D.G. \& Herrero R.C. Network analysis: An indispensable tool for curricula design. A real case-study of the degree on mathematics at the URJC in Spain. \textit{PLoS ONE} \textbf{16}(3), e0248208, (2021).
    
    \bibitem{Slim2014a}
    Slim A., Kozlick J., Heileman G.L., Wigdahl J. \& Abdallah C.T. Network analysis of university courses. \textit{Proceedings of the 23rd International Conference on World Wide Web}, 713–718, (2014).
    
    \bibitem{Aldrich2015}
    Aldrich P.R. The curriculum prerequisite network: modeling the curriculum as a complex system. \textit{Biochemistry and Molecular Biology Education} \textbf{43}(3), 168-180, (2015).
    
    \bibitem{Molontay2020}
     Molontay, R.,  Horv\'{a}th, N.,  Bergmann, J.,  Szekr\'{e}nyes D. \&  Szab\'{a}, M. Characterizing curriculum prerequisite networks by a student flow approach," \textit{IEEE Transactions on Learning Technologies}  \textbf{13}(3), 491-501, (2020)
     
     \bibitem{Newman2018}
     Newman, M.E.J. \textit{Networks: An Introduction.} Oxford University Press, Oxford, (2018).
     
     \bibitem{NBW2006}	
     Newman, M.E.J., Barab\'{a}si, A.-L. \& Watts, D.J. \textit{The Structure and Dynamics of
     Networks.} Princeton University Press, Princeton, (2006).
 
 	\bibitem{Dorogovtsev2010}
     Dorogovtsev, S.N. \textit{Lectures on Complex Networks.} Oxford University Press, Oxford, (2010).
 
     \bibitem{Easley2010}
     Easley, D. \& Kleinberg, J. \textit{Networks, Crowds, and Markets: Reasoning about a Highly Connected World.} Cambridge University Press, Cambridge, (2010).

	\bibitem{Gephi} 
	Gephi: The Open Graph Viz Platform,\\ 
	\url{https://gephi.org/}.
	
	\bibitem{GitHub}
	GitHub: A Repository with the Caltech CPN Data,\\ 
	\url{https://github.com/pstavrin/Course-Prerequisite-Networks}.
	
	\bibitem{BA1999}
	Barab\'{a}si, A.-L. \& Albert, R. Emergence of scaling in random networks. \textit{Science}
	\textbf{286}, 509-512 (1999).
	
	\bibitem{Broder2000}
	Broder, A., Kumar, R., Maghoul, F., Raghavan, P., Rajagopalan, S., Stata, R., Tomkins, A. \&  Wiener, J. Graph structure in the web. \textit{Computer Networks} \textbf{33}, 309-320 (2000).
	
	\bibitem{CSN2009}
	Clauset, A., Shalizi, C.R. \& Newman, M.E.J. Power-law distributions in empirical
	data. \textit{SIAM Review} \textbf{51}(4), 661-703 (2009).
	
	\bibitem{PageBrin1998}
	Page, L., Brin, S., Motwani, R. \& Winograd, T. The pagerank citation ranking: bringing order to the web. Technical Report, \textit{Stanford InfoLab} (1998).
	
	\bibitem{BrinPage1998}
	Brin, S. \& Page, L. The anatomy of a large-scale hypertextual web search engine. \textit{Computer Networks and ISDN Systems} \textbf{3}(1-7), 107-117 (1998).
	
	\bibitem{NetworkX}
	NetworkX: A Python Package for Network Analysis, \\
	\url{https://networkx.org/}.
	
	\bibitem{Anthonise1971}
	Anthonisse, J.M. The rush in a directed graph. Technical Report BN 9/71, \textit{Stichting Mathematisch Centrum}, Amsterdam (1971).
	
	\bibitem{Freeman1977}
	Freeman, L.C. A set of measures of centrality based upon betweenness. \textit{Sociometry} \textbf{40}, 35-41 (1977).
	
	\bibitem{Radicchi2009}
	Radicchi, F., Fortunato, S. \& Vespignani, A. Citation networks. In A. Scharnhorst, K. B\"{o}rner \& P. van den Besselaar, eds., \textit{Models of Science Dynamics: Encounters Between Complexity Theory and Information Science}, 233-257, Springer, New York (2012).
	
	\bibitem{Dunne2002}
	Dunne, J.A., Williams, R.J. \& Martinez, N.D. Food-web structure and network theory: The role of connectance and size. \textit{PNAS} \textbf{99}(20), 12917-12922 (2002).
	
	\bibitem{Bombelli1987}
	Bombelli, L., Lee, J., Meyer, D. \& Sorkin, R.D. Space-time as a causal set. \textit{Physical Review Letters} \textbf{59}(5), 521-524 (1987).
	
	\bibitem{Cowel1999}
	Cowel, R.G., Dawid, A.P., Lauritzen, S.L. \& Spiegelhalter, D.J.  \textit{Probabilistic
	Networks and Expert Systems.} Springer-Verlag New York (1999).

    \bibitem{Goodfellow2016}
    Goodfellow, I., Bengio, Y. \& Courville, A. \textit{Deep Learning}. MIT Press (2016).
    
    \bibitem{DAG}
    Sedgewick, R. \& Wayne, K. \textit{Algorithms (Fourth Edition)}. Addison-Wesley (2011).
	
	\bibitem{Zinoviev2018}
	Zinoviev, D. \textit{Complex Network Analysis in Python}. The Pragmatic Bookshelf (2018).
	
	\bibitem{GrivanNewman2002}
	Girvan, M. \& Newman, M.E.J. Community structure in social and biological networks. \textit{Proceedings of the National Academy of Sciences} \textbf{99}(12), 7821-7826 (2002).
	
	\bibitem{Porter2009}
	Porter, M.A., Onnela, J.-P. \& Mucha, P.J. Communities in networks. \textit{Notices of the American Mathematical Society} \textbf{56}(9), 1082-1097 (2009).
	
	\bibitem{Fortunato2010}
	Fortunato, S. Community detection in graphs. \textit{Physics Reports} \textbf{486}(3), 75-174 (2010).
	
	\bibitem{Holme2005}
	Holme, P. Core-periphery organization of complex networks. \textit{Phys. Rev. E} \textbf{72}(4), 046111 (2005).
	
	\bibitem{Csermely2013}
	Csermely, P., London, A., Wu, L.-Y. \& Uzzi, B. Structure and dynamics of core/periphery networks. \textit{Journal of Complex Networks} \textbf{1}(2), 93-123 (2013).
	
	\bibitem{Rombach2014}
	Rombach, M.P., Porter, M.A., Fowler, J.H. \& Mucha, P.J. Core-periphery structure in networks. \textit{SIAM Journal on Applied Mathematics} \textbf{74}(1), 167-190 (2014).
	
	\bibitem{Alvarez2005}
	Alvarez-Hamelin, J., Dall'Asta, L., Barrat, A. \& Vespignani, A. Large scale networks fingerprinting and visualization using the k-core decomposition. In Y. Weiss, B. Sch\"{o}lkopf \& J. Platt, eds., \textit{Advances in Neural Information Processing Systems}, vol. 18 (2005).
	
	\bibitem{Dufresne2016}
	H\'{e}bert-Dufresne, L., Grochow, J. \& Allard, A. Multi-scale structure and topological anomaly detection via a new network statistic: The onion decomposition. \textit{Scientific Reports} \textbf{6}, 31708 (2016).
	
	\bibitem{toposort}
	Toposort: A Python module that implements topological stratification of directed acyclic graphs,\\
	\url{https://pypi.org/project/toposort/}.
	
	\bibitem{CaltechCatalog}
	Caltech Catalog 2021-22, \url{https://catalog.caltech.edu/archive/2021-22/}. 
	
	\bibitem{MacKay}
	MacKay, D.J.C. \textit{Information Theory, Inference and Learning Algorithms}. Cambridge University Press (2002).
	
\end{thebibliography}
\end{document}